\documentclass{article}

\usepackage{arxiv}

\usepackage[utf8]{inputenc} 
\usepackage[T1]{fontenc}    
\usepackage{hyperref}       
\usepackage{url}            
\usepackage{booktabs}       
\usepackage{amsfonts}       
\usepackage{nicefrac}       
\usepackage{microtype}      

\usepackage[round]{natbib}
\usepackage{float}
\usepackage{graphicx}
\usepackage{subfigure}
\usepackage[font=small, labelfont=bf]{caption}
\usepackage{amsthm,amsmath,amsfonts}
\usepackage{multirow}
\usepackage{makecell}
\usepackage{longtable}

\title{Shape Matters: Evidence from Machine Learning on Body Shape-Income Relationship\thanks{The authors would like to express their appreciation to Petri B\"{o}ckerman, Gordon B. Dahl, David Frisvold, Dan Hamermesh, M. Ali Khan, Dan Silverman, and Harald Uhlig for numerous helpful suggestions. This paper has also benefited from the reactions of  seminar audiences at Microsoft Research in Redmond, Federal Reserve Bank at Kansas City, Georgia Institute of Technology, New York University, as well as from participants
at Conference on New Frontiers in Econometrics at Stamford, Joint Statistical Meetings at Vancouver, Midwest Econometrics Group at University of Wisconsin at Madison, Annual Conference of the Southern Economic Association at Washington D.C., and Econometrics Mini-conference at University of Iowa. 
}}

\author{
    Suyong Song\thanks{Corresponding author}\\
    Department of Economics\\
    University of Iowa\\
    Iowa City, IA 52242\\
    \texttt{suyong-song@uiowa.edu} \\
    \And
    Stephen Baek\\
    Department of Industrial and Systems Engineering\\ University of Iowa\\
    Iowa City, IA 52242\\
    \texttt{stephen-baek@uiowa.edu}\\
}

\begin{document}
\maketitle

\begin{abstract}
We study the association between physical appearance and family income using a novel data which has 3-dimensional body scans to mitigate the issue of reporting errors and measurement errors observed in most previous studies. 
We apply machine learning to obtain intrinsic features consisting of human body and take into account a possible issue of endogenous body shapes. The estimation results show that there is a significant relationship between physical appearance and family income and the associations are different across the gender. This supports the hypothesis on the physical attractiveness premium and its heterogeneity across the gender. 

\end{abstract}


\section{Introduction}

This paper studies the relationship between physical appearance and family income using unique three-dimensional (3D) whole-body scan data. Recent development in machine learning is adapted to extract intrinsic body features from the scanned data. Our approach underscores the importance of reporting errors and measurement errors on conventional measurements of body shape such as height and body mass index (BMI).

In the literature on the association between physical attractiveness and the labor market outcomes, sparse measurements such as facial attractiveness, height and BMI are mainly considered as measurements of the physical appearance. For instance, \citet{HamermeshBiddle94} studied the impact of facial attractiveness on wages and showed that there is significant beauty premium. \citet{PersicoPostlewaiteSilverman04} and \citet{CasePaxson08} analyzed the effects of height on wages. They found apparent height premium in the labor market outcomes.
\citet{Cawley04} estimated the effects of BMI on wages and reported that weight lowers the wages of white females.
In most previous studies, physical appearance was measured by imperfect proxies from subjective opinion based on surveys. This concerns a possibility of attenuation bias from reporting errors on the physical appearance in the estimation of the relation between physical appearance and labor market outcomes. In addition, measurements such as height, weight, and BMI are too sparse to characterize detailed body shapes. As a result, the issue of the measurement errors on the body shapes would make it difficult to correctly estimate the true relation so that many studies ended up with mixed results. 

We use a novel dataset, the Civilian American European Surface Anthropometry Resource (CAESAR) dataset. The dataset contains detailed demographics of subjects and $40$ anthropometric measurements such as height and weight, obtained using tape measures and calipers. It also contains the height and weight reported by subjects. This allows us to calculate the reporting errors in height and weight for each subject and investigate their properties and impacts on the estimation results. We found that the reporting error in males' height is correlated with their characteristics such as family income, age, race and birth region, while the reporting error in females' height is correlated with their age. We also found that the reporting error in males' weight is dependent with  their true weight, while the reporting error in females' weight is associated with their true weight and hours of exercises. 

We further investigate properties of reporting errors using a nonparametric conditional mean and nonlinear quantile functions. The nonparametric estimation of the conditional expectations of the reporting errors in height given the true height shows that the reporting error for female’s height is nonclassical in the sense that the reporting error and the true height are dependent. The quantile regression provides that the conditional median of the reporting error is independent of the true height. Thus it would be more plausible to impose a restriction on the conditional quantile of the reporting error of height than the conditional mean (see \citet{Bollinger98}, \citet{HuSchennach08}, and \citet{Song15}). On the other hand, the nonparametric conditional mean and nonlinear quantile regressions show that there are substantial nonclassical errors in both genders' reported weight.

The estimation results for the association between height (or BMI) and the family income confirm that the reporting errors have substantial impacts on the estimated coefficients. Furthermore, such conventional measurements on body shape are too sparse to describe whole body structure. So the analyses with the sparse measurements are very sensitive to the variable selection, which implies that regressions with the measured height and BMI might suffer from the issue of the measurement errors on the body shape. A handful of papers address the issue by proposing statistical methods such as bias-correction methods or instrumental-variables approaches which deliver consistent estimators at the expense of strong assumptions.

The dataset encloses digital 3D whole-body scans of subjects, which is a very unique feature. The scanned data on human body shapes would mitigate the issue of possible measurement errors due to the sparse measurements. Since the observed variable on body shapes in the dataset is three-dimensional, nevertheless, it is not straightforward to incorporate the data into the model of the family income. Indeed, there are 45,534 covariates for each individual's body scan.\footnote{A 3D whole-body scan for each subject contains $15,178$ number of vertices/nodes and each vertex consists of geometric $(x,y,z)$ coordinate. Thus there are $15,178\times 3=45,534$ inputs which is high dimensional.} To this end, we adopt methods based on machine learning to identify important features from 3D body scan data. Autoencoders are a certain type of artificial neural networks that possesses an hour-glass shaped network architecture. They are useful in extracting the intrinsic information from the high dimensional input and in finding the most effective way of compressing such information into the lower dimensional encoding. As shown in this paper, the graphical autoencoder can effectively extract the body features and is not sensitive to random noises. 
To the best of our knowledge, we are not aware of any studies published in Economics which use three-dimensional graphical data with such high dimensionality.

There have been increasing attentions to geometric data such as human body shapes, social networks, firm networks, product reviews, geographical models, etc, in economic studies. In this paper, we introduce new methodology built on deep neural networks and show how it can be utilized to analyze the economic model when the available data has a geometric structure.
When  one  attempts  to  incorporate geometric data in statistical analyses, there is no trivial grid-like representation for the data. As a result, encoding the features and characteristics of each data point into a numerical form is neither straightforward nor consistent. Most existing studies simplify the geometric features with some sparse characteristics. For instance, in the human body data, 
many of the relevant studies quantify the geometric characteristics of a human body shape with some sparse measurements, such as height and weight. However, such methods do not always capture detailed geometric variations and often lead to an incorrect statistical conclusion due to the measurement errors.
As a better alternative, we propose a graphical autoencoder that can interface with the three-dimensional graphical data. The graphical autoencoder permits incorporation of geometric manifold data into economic analyses. As we will discuss, direct incorporation of the graphical data can reduce measurement errors because graphical data in general provides more comprehensive information on geometric data compared to discrete geometric measurements.



From the proposed method using the graphical autoencoder, we successfully identify intrinsic features of the body shape from 3D body scan data.
Interestingly, intrinsic features of the body type are significantly important to explain the family income. Using the graphical autoencoder, we identify two intrinsic features forming male's body type and three intrinsic features for female's body type. In contrast to the conventional principle component analysis, the graphical autoencoder renders us to interpret the extracted features. For both genders, the first feature captures how tall a person is (stature), while the second feature captures how obese the body type is (obesity). The third feature captures hip-to-waist ratio of the body shape among the female sample.  

As acknowledged in the literature, body types could be endogenous in that these can be also driven by unobserved factors of income such as nutrition, personality, ability, and family background. In order to identify the causal impact of body types on family income, we correct for possible endogeneity issues of body types. We utilize proxy variables approach and control functions approach. In particular, our identification strategy is to use variations in shoe size, jacket size for males (blouse size for females), and pants size as legitimate instrumental variables for stature in the control functions approach. By testing the null of exogenous stature, we find that female's stature is endogenous but not male's.

\begin{figure}[t]
    \centering
    \includegraphics[width=\linewidth]{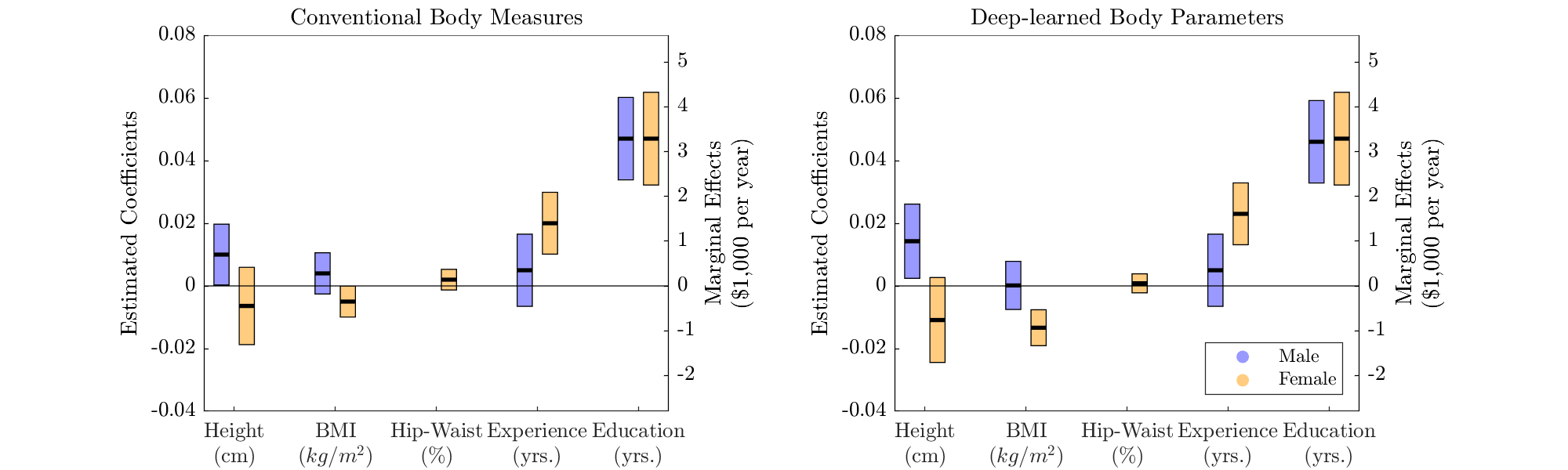}
    \caption{\textbf{Summary of the Estimation Results for Family Income Equation.} Estimated coefficients and bootstrapped 90\% confidence bands are reported. The left panel presents results from the conventional body measures and the right panel reports results from the deep-learned body parameters through the graphical autoencoder.}
    \label{fig:variables_boxplot}
\end{figure}

We summarize the main findings in Figure~\ref{fig:variables_boxplot}. In the estimation results from the deep-learned body parameters (right panel), we find that for males stature  has a positive impact on  family income and is statistically significant at 5\% significance level, while obesity  is insignificant. We estimate one centimeter increase in stature (converted in height) is  associated with approximately $\$998$ increase in the family income for a male who earns $\$70,000$ of median family income.
For females, the  coefficient of obesity is negative and statistically significant at 1\% significance level. On the other hand, coefficients of other features such as height and hip-to-waist ratio are statistically insignificant. One unit decrease in obesity (converted in BMI) is associated with approximately $\$934$ increase in the family income for a female who earns $\$70,000$ family income. The results imply there still exist physical attractiveness premium and its heterogeneity across the gender in the relationship between body shapes and income, even after controlling for unobserved confounding factors. Education is statistically significant for both genders but experience is significant only for the female samples.  However, in the estimation results from the conventional body measures such as height and BMI (left panel),  the magnitude of the estimated coefficients are much smaller than those from the deep-learned body parameters, which supports the possibility of attenuation bias due to measurement errors.
This could suggest that height and BMI have  limited powers to describe body shapes so any statistical analysis based on those simple measurements would lead to wrong economic inferences. Our findings also highlight the importance of correctly measuring body shapes to provide adequate public policies for the healthcare.

The rest of the article is organized as follows. Section 2 presents the model of interest. Section 3 introduces and summarizes the CAESAR dataset. Section 4 discusses the impact of reporting errors in height and weight.
Section 5 discusses estimation results for the impact of the physical appearance on family income. Section 6 concludes. Empirical estimation results are contained in Appendix.

\section{Model}

We consider the association between family income and body shapes as follows:
\begin{equation} \label{eq:income-body}
\text{Family Income}_{i}=\alpha X_{i} + \beta\text{Body Shapes}_{i} + \epsilon_{i}, \hspace{5mm} i=1,..., N,
\end{equation}
where $\text{Family Income}_{i}$ is log family income,  $\text{Body Shapes}_{i}$ is a measure of body types, $X_{i}$ is a set of covariates, and $\epsilon_{i}$ is unobserved causes of family income for individual $i$. We are particularly interested in the parameter $\beta$, but we also discuss the relationship between family income and other individual characteristics through the vector of parameters $\alpha$. We consider family income instead of individual income, since family income is available in the data. So we identify the combined effects of body shapes on income through the labor market and marriage market. In fact, as documented in \citet{ChiapporiOrefficeQuintana12}, various studies found assortative matching on income, wages, education, and anthropometric characteristics such as weight or height in marriage market. As a result, total effects of body shapes in the labor market and marriage market are identifiable from family income so that it is worthwhile to investigate the impact of physical attractiveness on family income.

A large body of literature has analyzed the presence of earnings differentials based on physical appearance. A strand of literature has focused on facial attractiveness. \citet{HamermeshBiddle94} analyzed the effect of physical appearance on earnings using interviewers' ratings of respondents' physical appearance. They found evidence of a positive impact of looks on earnings. \citet{MobiusRosenblat06} examined the sources of the beauty premium and decomposed the beauty premium that arises during the wage negotiation process between employer and employee in an experimental labor market.
They identified transmission channels through which physical attractiveness raises an employer's estimate of a worker's ability. \citet{ScholzSicinski15} studied the impact of facial attractiveness on the lifetime earnings. They found there exists the beauty premium even after controlling for other factors which enhance productivity in the labor market earnings.

Other threads of literature have analyzed the effects of height on labor market outcomes. \citet{PersicoPostlewaiteSilverman04} found 
that an additional inch of height is associated with an increase in wages, which is a consistent finding in the literature in addition to racial and gender bias. They showed that how tall a person is as a teenager is the source of the height wage premium. This implies that there are positive effects of social factors associated with the accumulation of productive skills and attributes on the development of human capital and the distribution of economic outcomes.
\citet{CasePaxson08} also found there are substantial returns to height in the labor market. However, they showed that the height premium is the result of positive correlation between height and cognitive ability.
\citet{LundborgNystedtRooth14} found that the positive height-earnings association is explained by both cognitive and noncognitive skills observed in tall people.
\citet{DeatonArora09} reported that taller people evaluate their lives more favorably and the findings are explained by the positive association between height and both family income and education.
\citet{BockermanVainiomaki13} used twin data to control for unobserved ability and found a significant height premium in wage for women but not for men.
\citet{Lindqvist12} studied the relationship between height and leadership and confirmed that tall men are significantly more likely to attain managerial positions.

\citet{Cawley04} considered the effects of obesity on wages. He found that weight lowers the wages of white females and noted that one possible reason for the result is that obesity has adverse impact on the self-esteem of white females.
\citet{Rooth12} used a field experimental approach to find differential treatment against obese applicants in terms of the number of callbacks for a job interview in the hiring process in the Swedish labor market. He found the callback rate to interview was lower for both obese male and female applicants than for nonobese applicants. 

Mathematically, human body shapes can be viewed as arbitrary 2-manifolds $\mathcal{M}$ embedded in the Euclidean 3-space $\mathbb{R}^3$. In statistical analyses as in equation (\ref{eq:income-body}), quantifying geometric characteristics of different manifold shapes and encoding them into a numerical form is not straightforward. Thus, these continuous manifolds are approximated by proxies in a tensor form.
Due to this reason, many of the relevant works in the literature on the physical appearance  quantify the geometric characteristics of a human body shape with some sparse measurements, such as height, weight, or BMI. As we will see in the later sections, however, such kind of quantification methods do not always capture detailed geometric variations and often lead to an erroneous explanation of statistical data. For instance, with height and BMI alone, one can hardly distinguish muscular individuals from individuals with round body shapes. The situation does not improve even if some new variables, such as chest circumference, are added, since these variables still are not quite enough to codify all the subtle variations in body shapes. Moreover, oftentimes, such additional variables merely add redundancy, without adding any substantial statistical description of data, as the commonly-used anthropometric parameters are highly correlated to each other. In addition, it is also noteworthy that the manual selection of measurement variables can also introduce one's bias into the model.
In this paper, we compare several common ways of quantifying manifold structured data with a newly-proposed graphical autoencoder method.

\section{Data}

We use a unique data, called the Civilian American European Surface Anthropometry Resource (CAESAR) dataset. It was collected from a survey of the civilian populations of three countries representing the North Atlantic Treaty Organization (NATO) countries: the U.S., The Netherlands, and Italy. The survey was primarily conducted by the U.S. Air Force and the 
sample from the U.S. was used for our study. The survey of the U.S. sample was conducted from 1998 to 2000 and carried out in 12 different locations which were selected to obtain subjects approximately in proportion to the proportion of the population in each of 4 regions of the U.S. Census.\footnote{The U.S. data is referred to as the North American sample since one site in Ottawa, Canada was added to the sample.}

The dataset contains 2,383 individuals whose ages vary from 18 to 65 with a diverse demographical population. The dataset contains detailed demographics of subjects, anthropometric measurements done with a tape measure and caliper, and digital 3D whole-body scans of subjects.
In contrast to other traditional surveys, 
the data contains both reported and measured height and weight. This feature makes it possible to calculate reporting errors in the survey data and analyze their relations to the correctly measured height/weight as well as individual characteristics. In addition, the existence of 3D whole-body scan data makes the CAESAR data serve as a good proxy to physical appearance such that potential issue of measurement errors can be mitigated.

Some of the total 2,383 subjects in the database have missing demographic and anthropometric information; these have been deleted in our study. In addition, there are also subjects who elected not to disclose and/or were not aware of their income, race, education, etc. These individuals have also been removed in this study. In the analysis, we divide the sample by gender to take into account the differential treatment across genders. 

Tables \ref{tb:summarystat-men}-\ref{tb:summarystat-women} provide summary statistics of the variables in the database for males and females, respectively. The data has a single question about family income (grouped into ten classes). Average family income is \$76,085 for males and \$65,998 for females. The differences in average family income across genders would be due to the fact that the male sample includes more married people than the female sample. Median family income is slightly lower than the mean family income, which amounts to \$70,000 for males and \$52,500 for females. 
For males, on average, reported height is 179.82 centimeters and measured height is 178.26 centimeters, which shows a tendency of over-reporting. The gap is larger when median reported height (180.34 centimeters) and measured height (177.85 centimeters) are compared. We observe a similar pattern in the female sample: reported height is 164.96 centimeters and measured height is 164.22 centimeters on average; median reported height is 165.1 centimeter and median measured height is 164 centimeters. 

The males' average reported weight is 86.03 kilograms and the average of the measured weight is 86.76 kilograms. The median of two measurements are the same. For females, reported weight is 67.88 kilograms and measured weight is 68.81 kilograms on average. Median reported weight is 63.49 kilograms and median measured weight is 64.85 kilograms. In both subsamples, the standard errors of the weight are large, which are approximately 17 kilograms.
BMI has been commonly used as a screening tool for determining whether a person is overweight or obese.\footnote{According to Centers for Disease Control and Prevention (CDC), the standard weight status categories associated with BMI rannges for adults are as follows: below 18.5 (underweight), 18.5-24.9 (normal or healthy weight), 25.0-29.9 (overweight), 30.0 and above (obese).} BMI is calculated as weight in kilograms divided by height in meters squared. We refer reported BMI (measured BMI) to the one based on reported height and weight (measured height and weight). In the tables, height, weight and BMI are those measured by professional tailors at the survey sites. For both genders, reported BMI is slightly larger than measured BMI on average.

In addition to the bio-metric measurements, the data contains other variables for individual characteristics and socio-economic backgrounds. Education grouped into nine categories is 16.29 years for males and 15.75 years for females on average. Experience is calculated as potential $\text{experience}=\text{age}-\text{education}-6$ and its mean is  17.54 years for males and 18.62 years for females. Fitness is defined as exercise hours per week. Its mean and median are 4.24 hours and 2.5 hours, respectively, for males. For females,  its mean and median are 3.74 hours and 2.5 hours, respectively.

The data also include the number of children.  Marital status is classified as three groups: single, married, divorced/widowed. Occupation consists of white collar, management, blue collar, and service. Race has four groups including White, Hispanic, Black, and Asian. Birth region is grouped into five groups including Midwest, Northeast, South, West, and Foreign.
The majority in the dataset are white collar married White males and females born in Midwest. As we will discuss later, the data also contains 40 body measures which includes height and weight. The list of the body measures are provided in Table \ref{tb:list-of-bodymeasures}.



\section{Reporting Errors in Height and Weight}

Several studies in the literature use survey data so that they assume there are no reporting errors in height and weight or reporting errors are classical in that they are not correlated with true measures.\footnote{Exceptionally, \citet{PersicoPostlewaiteSilverman04}  and \citet{CasePaxson08} use measured height from the British National Child Development Survey, even though they also use self-reported height from the  British Cohort Study and the National Longitudinal Survey of Youth, respectively. \citet{LundborgNystedtRooth14} use measured height from the Swedish National Service Administration.}
Since our data contains both reported and measured height and weight, we can further investigate the properties of the reporting errors. We consider measured height and weight as the true height and weight since they are measured by professional tailors. The reporting errors are calculated as $\text{Reporting Error}^{H}=\text{Reported Height} - \text{Height}$ and $\text{Reporting Error}^{W}=\text{Reported Weight}- \text{Weight}$, respectively

\begin{figure}[t]
     \begin{centering}
     \includegraphics[width=\linewidth]{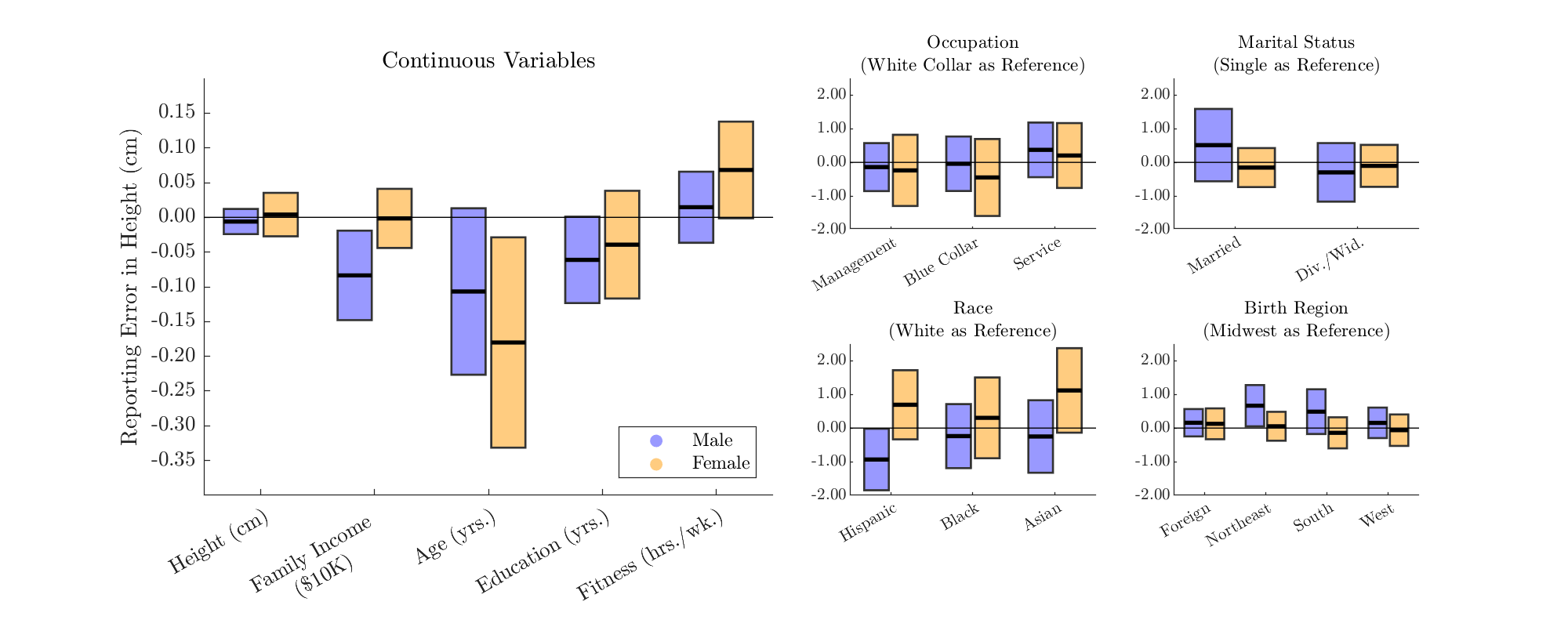}
     \end{centering}
     \vspace*{-0.2cm}
     \caption{\textbf{Personal Background and Reporting Errors in Height.}
     Estimated coefficients and bootstrapped 90\% confidence bands are reported. Note that the unit for height is converted into centimeter (cm).}
     \label{fig:table4_height}
 \end{figure}

The following equation estimates which personal background explains reporting error in height and weight:
\begin{align} \label{eq:ReportingError-Height}
\text{Reporting Error}_{i}^{H}&=\pi X_{i} +\mu \text{Height}_{i}  + \varepsilon_{i} ,\\
\label{eq:ReportingError-Weight}
\text{Reporting Error}_{i}^{W}&=\pi X_{i} +\mu \text{Weight}_{i} + \varepsilon_{i} ,
\end{align}
where $X_i$ is a set of covariates including  family income, age, age squared, occupation, education, marital status, fitness, race, and birth region.  $\text{Height}_{i}$ is the true height in millimeters, and $\text{Weight}_{i}$ is the true weight in kilograms.
We found dependence between reporting errors and some covariates. Table \ref{tb:reporting-error} reports the estimation results. The standard errors are estimated by bootstrapping and are reported inside the parentheses. In the equation (\ref{eq:ReportingError-Height}), the coefficient of the true height is not statistically significant for both genders.  We observe different results across the gender. For males, family income is negatively correlated with the reporting error in height at 1\% significance level. Age squared is positively correlated with the reporting error.
Hispanic males are more likely to under-report their height compared to White males. Males who were born in Northeast are more likely to over-report their height relative to those from Midwest. On the other hand, the coefficient of family income is not statistically significant for females. Older females are more likely to under-report their height. The estimation results are summarized in Figure \ref{fig:table4_height}.

\begin{figure}[t]
     \begin{centering}
     \includegraphics[width=\linewidth]{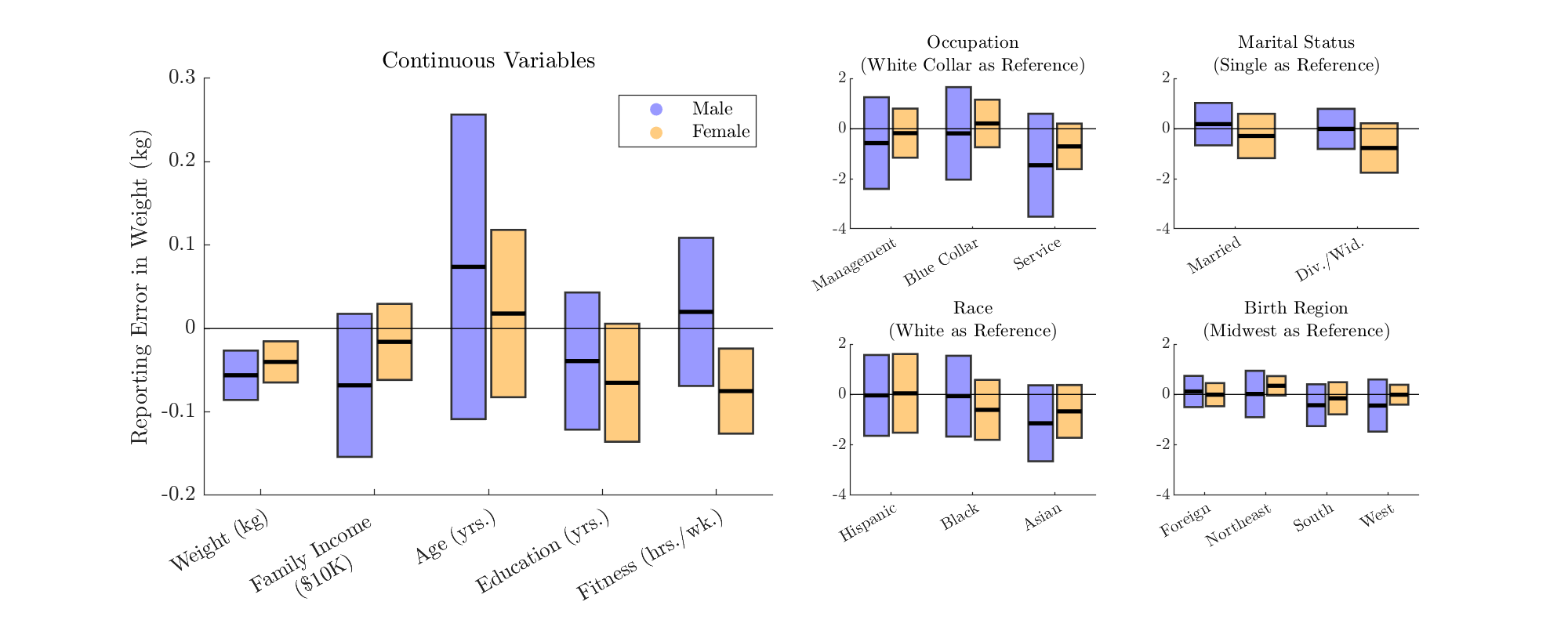}
     \end{centering}
     \vspace*{-0.2cm}
     \caption{\textbf{Personal Background and Reporting Errors in Weight.} Estimated coefficients and bootstrapped 90\% confidence bands are reported.}
     \label{fig:table4_weight}
 \end{figure}

In the equation (\ref{eq:ReportingError-Weight}), the true weight is negatively correlated with the reporting error in weight (at 1\% significance level) for both genders: heavier people have a tendency to under-report their weight. For females, it is interesting to find that  the coefficient of fitness is statistically significant at 5\% significance level and it is negatively correlated with reporting-error in weight. Thus females who spend more time on exercise have a tendency to under-report their weight. However, we find little evidence that other personal background are correlated with reporting error in weight. The estimation results are summarized in Figure \ref{fig:table4_weight}.

\begin{figure}[t]
    \begin{centering}
    \includegraphics[width=\linewidth]{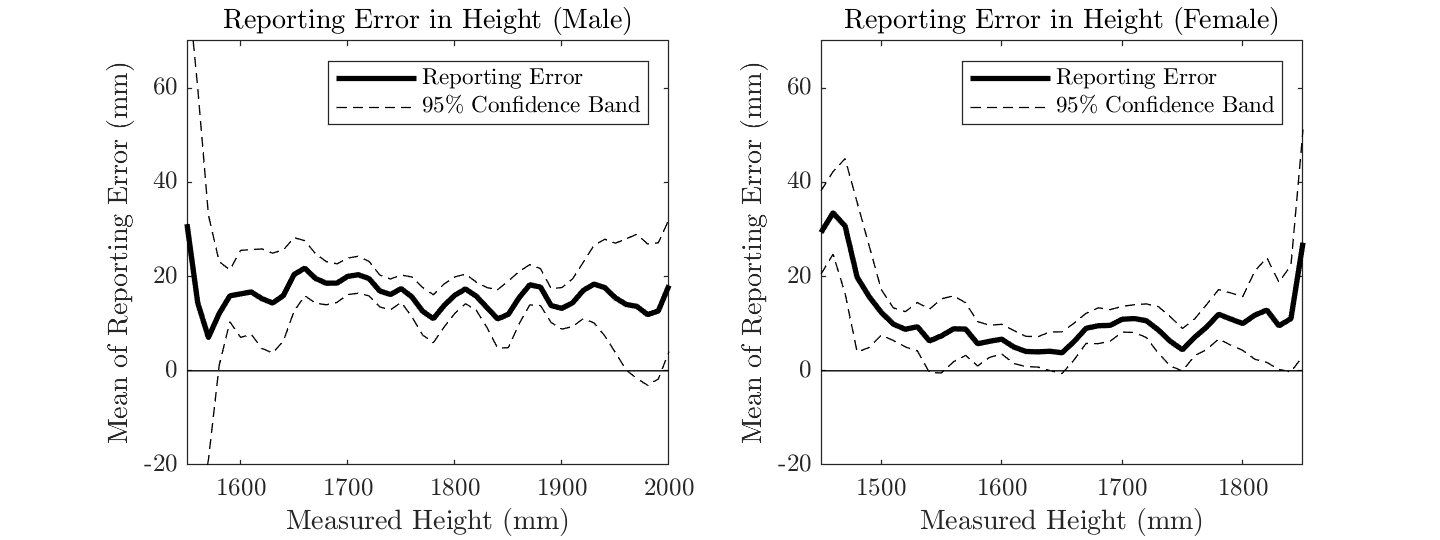}
    \end{centering}
    \vspace*{-0.3cm}
    \caption{\textbf{Conditional Mean of Reporting Errors in Height Conditional on True Height.}}
    \label{fig:ReportingError-TrueHeight}
\end{figure}

\begin{figure}[t]
    \begin{centering}
    \includegraphics[width=\linewidth]{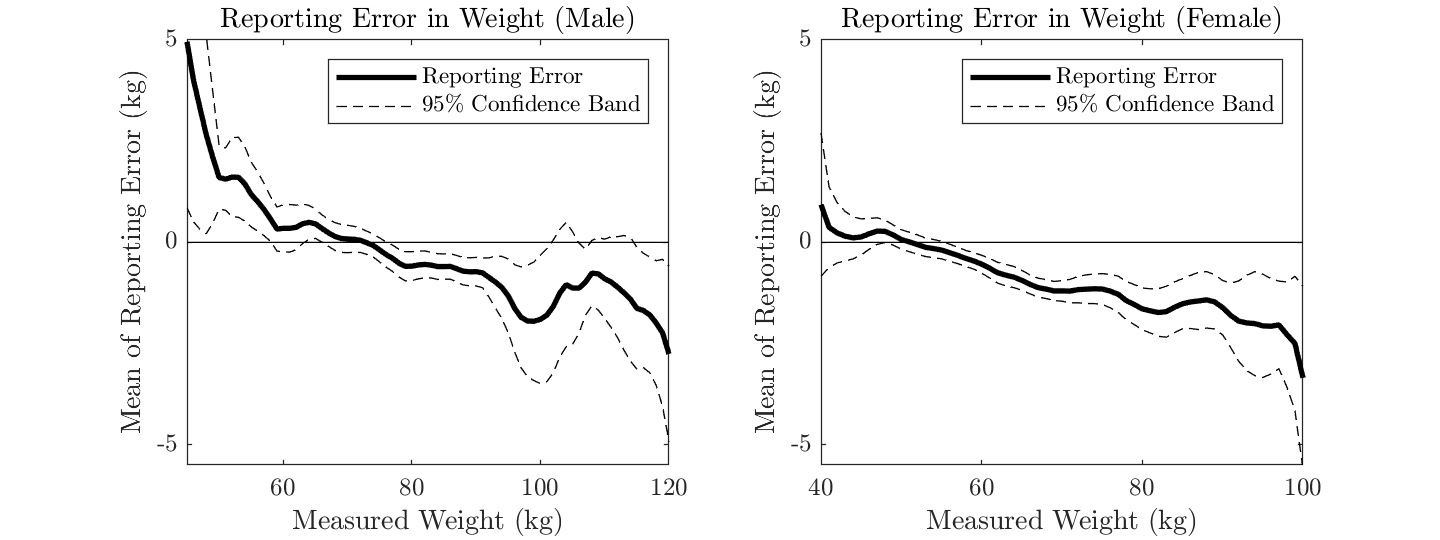}
    \end{centering}
    \vspace*{-0.3cm}
    \caption{\textbf{Conditional Mean of Reporting Errors in Weight Conditional on True Weight.}}
    \label{fig:ReportingError-TrueWeight}
\end{figure}

Figures \ref{fig:ReportingError-TrueHeight} and \ref{fig:ReportingError-TrueWeight} plot the  estimation of the conditional expectations of the reporting errors in height/weight given the true measures, namely, $E[\text{Reporting Error}^{H} \mid \text{Height}]$ and $E[\text{Reporting Error}^{W} \mid \text{Weight}]$ with their 95\% confidence bands, respectively. We use nonparametric kernel estimation where the kernel function is an  Epanechnikov kernel and the bandwidth is chosen by the Silverman's rule-of-thumb. The confidence bands are estimated by a nonparametric bootstrap method. The solid line represents zero reporting error. The nonparametric plots for the height show that the reported height is  larger than the true height at almost all height level in both genders showing over-reporting patterns. It displays no significant relation between the reporting error and the true height for males. For females, we observe more reporting error at low height level than at the average height level, which indicates that the reporting error for females' height is nonclassical in the sense that the reporting error and the true height are dependent.
This result is a finding which is not captured by the linear mean regression  in Table \ref{tb:reporting-error} where the reporting error in height is not related to the true height for both genders. 

The plots for the reported weight show more substantial nonclassical errors. Males who are at the low weight level (below approximately 75 kilogram) have a tendency to over-report their weight, but males at the weight level above approximately 75 kilogram under-report their weight. Similarly, females who are at the low weight level (below approximately 50 kilogram) have a tendency to over-report their weight, but  females at the weight level above approximately 50 kilogram under-report their weight. The both plots display an apparent dependence between the reporting error and the true weight.
This confirms the significant negative relation between the reporting error and the true weight shown in Table \ref{tb:reporting-error}.

\begin{figure}[t]
    \begin{centering}
    \includegraphics[width=\linewidth]{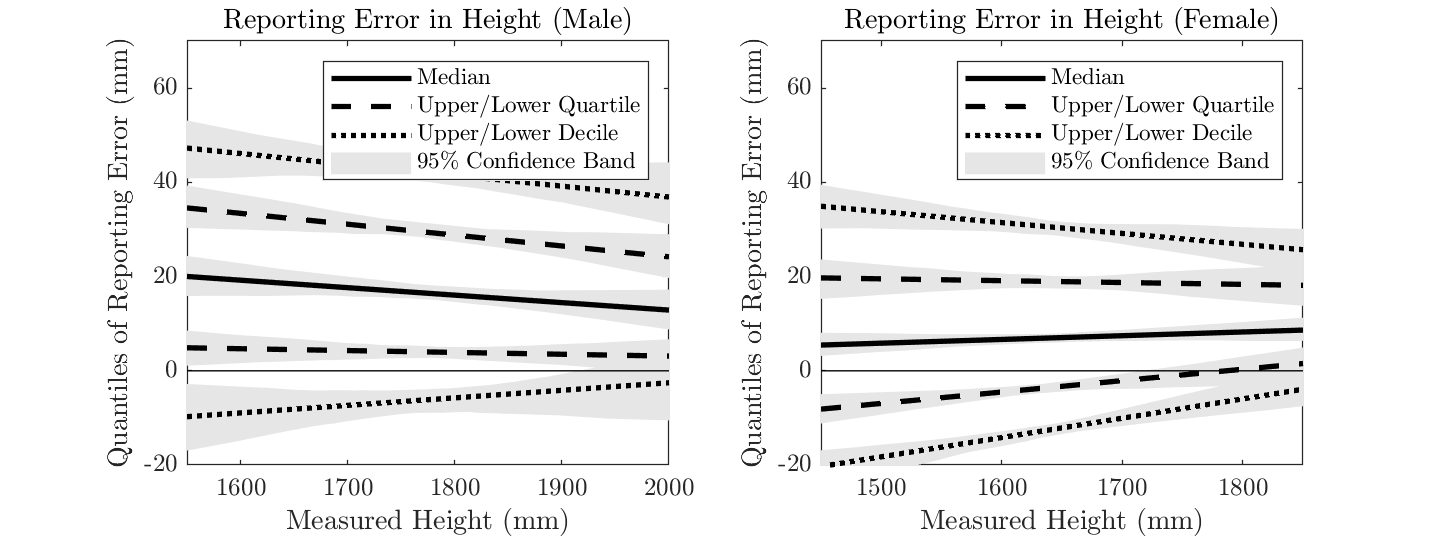}
    \end{centering}
    \vspace*{-0.3cm}
    \caption{\textbf{Conditional Quantile of Reporting Errors in Height Conditional on True Height.}}
    \label{fig:ReportingError-TrueHeight-Quantile}
\end{figure}

\begin{figure}[t]
    \begin{centering}
    \includegraphics[width=\linewidth]{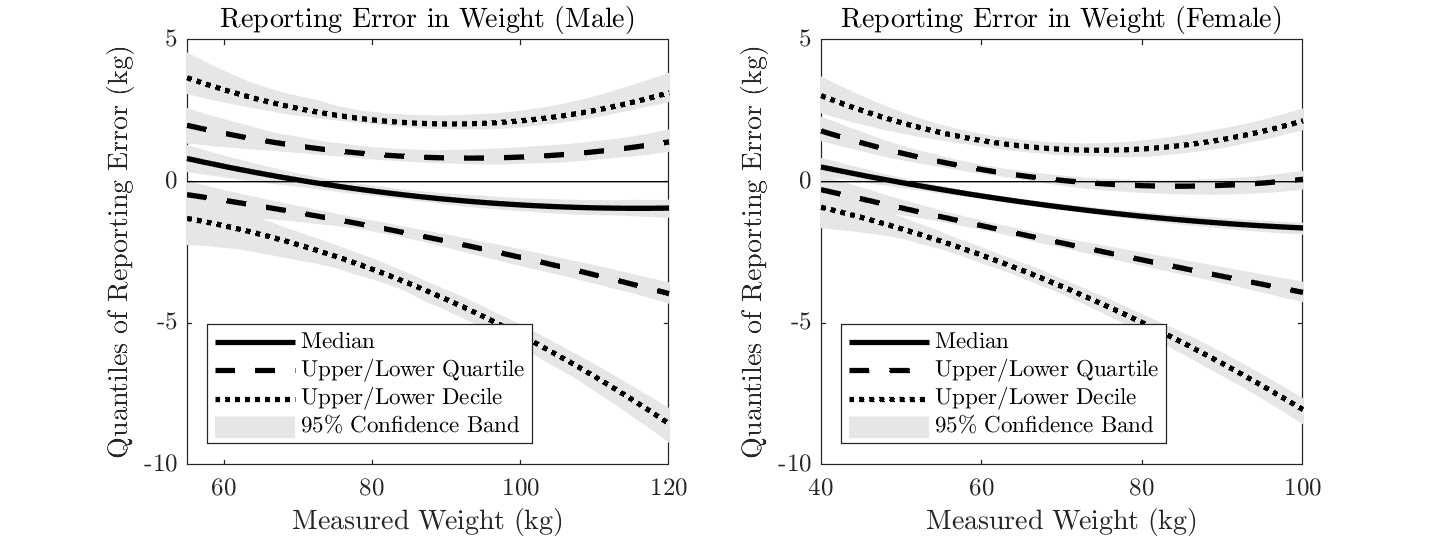}
    \end{centering}
    \vspace*{-0.3cm}
    \caption{\textbf{Conditional Quantile of Reporting Errors in Weight Conditional on True Weight.}}
    \label{fig:ReportingError-TrueWeight-Quantile}
\end{figure}

Conditional quantile function is a useful tool to estimate heterogeneity in a conditional distribution. It also measures what proportion of reporting errors are positive or negative. Figures \ref{fig:ReportingError-TrueHeight-Quantile}-\ref{fig:ReportingError-TrueWeight-Quantile} present the estimation of the conditional quantiles of the reporting errors in height and weight conditional on the true measures, namely, $Q_{\tau}[\text{Reporting Error}^{H} \mid \text{Height}]$ and $Q_{\tau}[\text{Reporting Error}^{W} \mid \text{Weight}]$ for $\tau \in (0,1)$ with their 95\% confidence bands, respectively. We estimate the conditional quantiles using the nonlinear polynomial regression. The figures display the median and the 10\%, 25\%, 75\%, and 90\% quantiles across genders. In Figure \ref{fig:ReportingError-TrueHeight-Quantile}, the results show that there is heterogeneity in the conditional distribution of the reporting error in height for both genders. It is shown that over-reporting of height is more pronounced for males than females. We notice that more than 75\% of the sample of the males over-report their height.  Interestingly, the median regression lines in both genders are approximately parallel with the horizontal line, which implies that the conditional median of the reporting error is independent of the true height. \citet{Bollinger98} also found that median regressions for earnings will be more robust to the reporting error than mean regressions. Thus, it would be more natural to impose a restriction on the conditional quantile of the reporting error of height than the conditional mean, as in \citet{HuSchennach08}.

Figure \ref{fig:ReportingError-TrueWeight-Quantile} also displays apparent heterogeneity in the conditional distribution for both genders. It shows that when heavier people than the average are concerned, under-reporting of weight is more pronounced for females than males. We notice that within this group, almost 75\% of the sample of the females under-report their weight.
The median regressions in both genders are dependent on the level of the measured weight. This indicates that there are substantial nonclassical errors in the reported weight so that a restriction on the conditional quantile may not be valid.

\section{Estimation of the Association between Physical Appearance and Labor Market Outcomes}

In this section, we estimate the association between the physical appearance and family income using various methods. 

\subsection{Height, Weight and Reporting Errors}


Most papers in the literature  estimate the relationship in the equation (\ref{eq:income-body}) by replacing body shapes with their observed proxies such as height or weight. However, these measurements are hardly accurate to fully describe body shapes. Furthermore, only including either height or weight without controlling for the other as in the literature could suffer from an omitted variable problem. For instance, consider there are two people who have the same height but different weight. Then comparing height only will fail to identify the difference in their body shapes.
Thus, we consider the following two regression equations:
\begin{align} \label{eq:Height1}
\text{Family Income}_{i}&=\alpha X_{i} + \beta_{1}\text{Height}_{i}  + \epsilon_{i},\\
\label{eq:Height2}
\text{Family Income}_{i}&=\alpha X_{i} + \beta_{1}\text{Height}_{i}  + \beta_{2}\text{Weight}_{i}+ \epsilon_{i},
\end{align}
where  $X_{i}$ is a set of controls including experience, $\text{experience}^2$, race, occupation, education, marital status, and number of children. We can test the importance of controlling for weight by comparing the estimated coefficients from two equations.
In addition, as mentioned before, the data contains measurements on height and weight both reported by subjects and measured by on-site measurers. Therefore, by comparing estimates of measured one with reported counterpart, we can see how much the reporting errors affect the estimation results. Table \ref{tb:reported-height/weight-income} reports estimation results from reported height and weight. Table \ref{tb:height/weight-income} provides estimation results from measured height and weight.

The hypothesis that the coefficient on height is zero is tested across gender. Results for both genders are presented in each tables. In equation (\ref{eq:Height1}) of Table \ref{tb:reported-height/weight-income}, reported weight is not included. The column for males shows that  education is statistically significant in  the income equation. The coefficient of the reported height is positive and statistically significant at 10\% significance level.
The column for females is somewhat different from that for males: the coefficient of experience, $\text{experience}^2$, and education  are statistically significant. In addition, the coefficient on the reported height is positive and statistically significant at 5\% significance level.
In equation (\ref{eq:Height2}), we add the reported weight to the set of regressors. The column for males shows that the coefficient of the reported height becomes statistically insignificant, and the coefficient of the reported weight is also insignificant. 
However, in the column for females the coefficient of the reported height is still positive and statistically significant, but the coefficient on the reported weight is insignificant.

In Table \ref{tb:height/weight-income}, we instead use the measured height and weight to estimate the income equation.
Interestingly, the coefficients on the height for males in both equations are positive and statistically significant at 1\% significance level. Their magnitudes are  larger than those from Table \ref{tb:reported-height/weight-income}. When the measured weight is added, its coefficient is still insignificant for males. For females, the coefficients on the measured height  are statistically significant in both equations and their magnitudes are larger than those from Table \ref{tb:reported-height/weight-income}. When the measured weight is added, its coefficient becomes significant at 10\% significance level, which shows a negative association between family income and weight. Thus, we confirm there are apparent reporting errors in height and weight. Particularly, the impacts of the reporting errors on the estimation results are more severe in males than females. These reporting errors bring attenuation bias to the estimates. Furthermore, 
the estimation results from two equations (\ref{eq:Height1})-(\ref{eq:Height2}) are different. 
It shows that using height only as a proxy to body shapes might be too simple to describe delicate figures of the physical appearance. 

As in \citet{Cawley04}, we consider BMI as the primary regressor in the regression equations (\ref{eq:Height1})-(\ref{eq:Height2}) to estimate the impact of obesity on income. Again, we add weight or height as additional regressor to take into account a possible omitted variable problem. So we consider the income equations as following:
\begin{align} \label{eq:BMI1}
\text{Family Income}_{i}&=\alpha X_{i} + \beta_{1}\text{BMI}_{i}  + \epsilon_{i},\\
\label{eq:BMI2}
\text{Family Income}_{i}&=\alpha X_{i} + \beta_{1}\text{BMI}_{i}  + \beta_{2}\text{Weight}_{i}+ \epsilon_{i},\\
\label{eq:BMI3}
\text{Family Income}_{i}&=\alpha X_{i} + \beta_{1}\text{BMI}_{i}  + \beta_{2}\text{Height}_{i}+ \epsilon_{i},
\end{align}
where $\text{BMI}_{i}$  is the body mass index. We first estimate the equations using the reported variables and summarize the estimation results in Table \ref{tb:reported-BMI-income}.
From the columns for males in the table, the coefficient of the reported BMI in equation (\ref{eq:BMI1}) is statistically insignificant. Adding the reported weight or height does not change the result for the reported BMI as in equations (\ref{eq:BMI2})-(\ref{eq:BMI3}). Instead, the estimated coefficient of the reported height or weight is significant.
For females, the coefficients of the reported BMI are insignificant in equations (\ref{eq:BMI1}) and (\ref{eq:BMI3}). However, in equation (\ref{eq:BMI2}), the reported BMI has a negative impact on family income and the relation is statistically significant at 1\% significance level. It also shows that the coefficient of the reported height or weight is positive and significant at 5\% significance level.

\begin{figure}[t]
     \centering
     \includegraphics[width=0.5\linewidth]{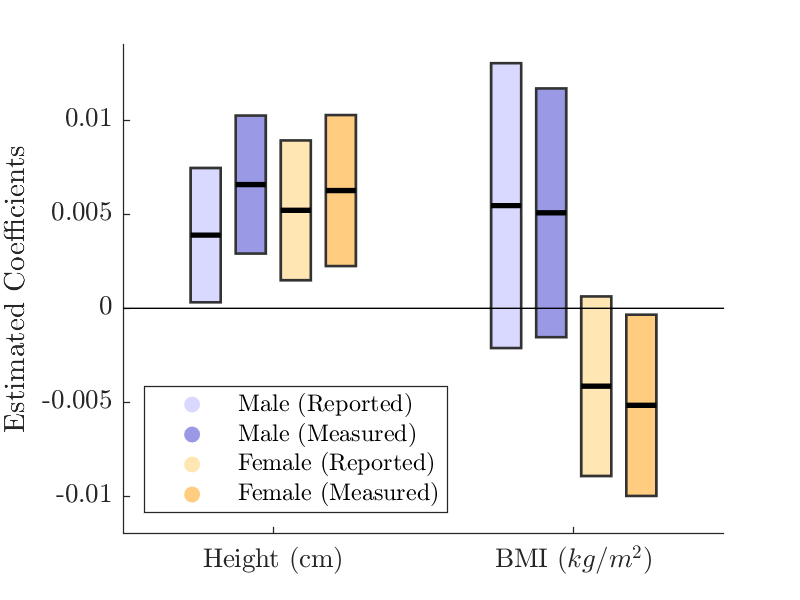}
     \vspace*{-0.3cm}
     \caption{\textbf{Comparison of Reported and Measured Body Measures.}
     Estimated coefficients and bootstrapped 90\% confidence bands are reported. We provide results from reported body measures (Reported) and measured body measures (Measured). Note that the unit for height is converted into centimeter (cm).}
     \label{fig:reported_vs_measured}
 \end{figure}

We next estimate equations (\ref{eq:BMI1})-(\ref{eq:BMI3}) using the measured BMI, height and weight. For all equations in Table \ref{tb:BMI-income}, the estimation results are somewhat different from Table \ref{tb:reported-BMI-income}.
In equation (\ref{eq:BMI1}) for males, the coefficient of BMI is still insignificant. When weight is included as in equation (\ref{eq:BMI2}), its coefficient for males is positive and statistically significant at 1\% significance level. The coefficient of BMI becomes negative and statistically significant at 5\% significance level. When height is included as in equation (\ref{eq:BMI3}), its coefficient for males is positive and statistically significant at 1\% significance level. However, the  coefficient of BMI is  statistically insignificant. For females, the results are different from those in Table \ref{tb:reported-BMI-income}. The coefficient of BMI is always negative and statistically significant. The coefficient of height or weight is positive and also statistically significant. The results for equation (\ref{eq:BMI3}) are highlighted in Figure \ref{fig:reported_vs_measured}. It is shown that the magnitudes of the coefficients for height are larger when measured height  are used. For females, the impact of the measured BMI on family income is significant  in contrast with insignificant impact of the reported BMI. Different signs on the effect of BMI  across genders are also observed: positive effect of male BMI as opposed to negative effect of female BMI. Thus, the analysis confirms that the reporting errors in body measures have significant impacts on the estimated coefficients.

Interestingly, we also observe that the estimation results have significantly changed as different set of measures of body types were included in the equations. 
One possible explanation for this is that even the measured height and BMI might not be perfect proxies to the body types, although they are less prone to reporting errors. In fact, height, weight and BMI are simple measures of body types so that they might miss useful information on the true body types (e.g., see \citet{WadaTekin10} for BMI).\footnote{Several studies propose statistical methods to reduce the measurement errors in body-shape measurements. Among others, \citet{CourtemanchePinkstonStewart15} propose a rank-based correction method for using validation data to correct the measurement errors in obesity. \citet{Murilloetal19} reduce bias in  obesity by applying regression calibration, simulation extrapolation, and multiple imputation approaches.}

In order to further investigate the role of the measurement errors
on the body types, we run the following regression equation:
\begin{align} \label{eq:part-of-body}
\text{Family Income}_{i}=\alpha X_{i} + \beta\text{Body}_{i} + \epsilon_{i},
\end{align}
where $\text{Body}_{i}$ is a set of body measurements which include 40 number of measurements on various parts of body.\footnote{A full list of the measurements is provided in Table \ref{tb:list-of-bodymeasures}.} Since these are more sophisticate than simple measurements of height and BMI, it is less likely that the measurement errors on body type is prevalent. 

\begin{figure}[t]
     \subfigure{
     \includegraphics[width=0.63\linewidth]{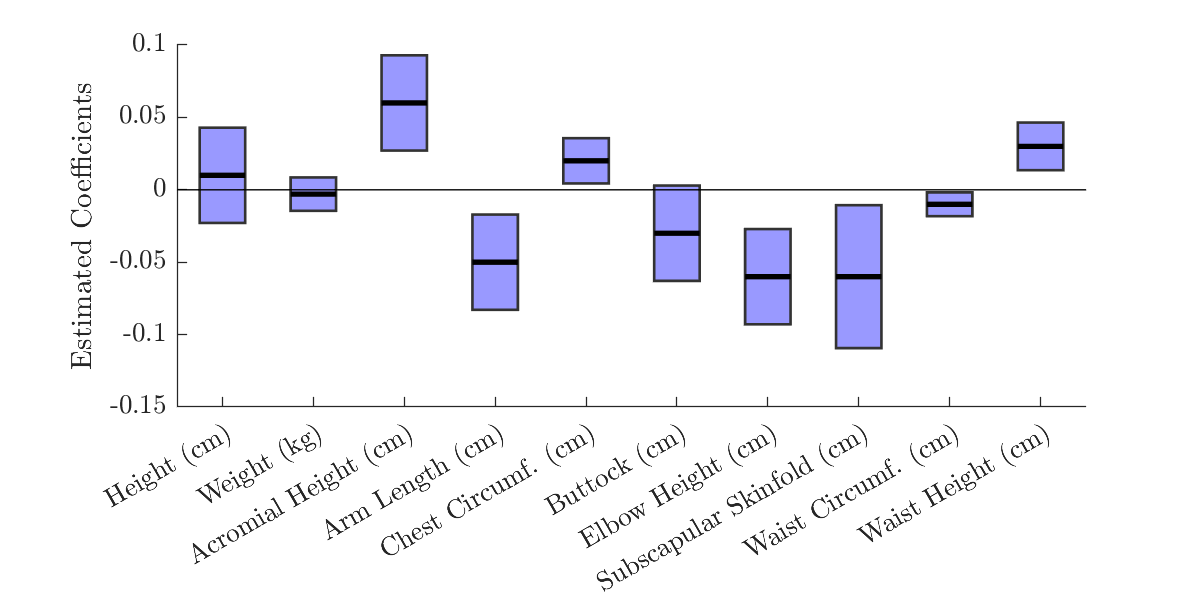}}
     \subfigure{
     \includegraphics[width=0.37\linewidth]{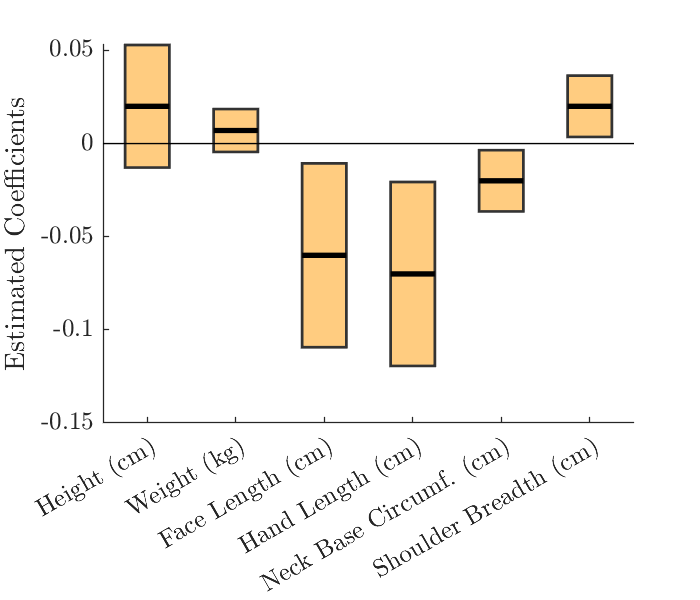}}
     \vspace*{-0.3cm}
     \caption{\textbf{Various Measures and Family Income.}
     Estimated coefficients and bootstrapped 90\% confidence bands are reported for male (\textit{left}) and female (\textit{right}). Note that units for all measurements  except weight are converted into centimeter (cm).}
     \label{fig:various_measures_income}
 \end{figure}

Table \ref{tb:bodymeasures-income} presents the estimation results. Except height and weight, for brevity, we only report measures of body parts which are statistically significant. Coefficients on age, race, occupation, education, and marital status are very similar to those in Table \ref{tb:BMI-income} for both genders. Interestingly, we found eight  statistically-significant body measurements for males and four for females (see Figure \ref{fig:various_measures_income}). For instance, in the sample of males, Acromial Height (Sitting), Chest Circumference, and Waist Height (Preferred)  have positive association with the family income, while Arm Length (Shoulder-to-Elbow), Buttock (Knee Length), Elbow Height (Sitting), Subscapular Skinfold, Waist Circumference (Preferred) are negatively correlated with the family income. For females, Shoulder Breadth is positively correlated with the family income. However, the coefficients on Face Length, Hand Length, Neck Base Circumference are all negative. 
The most distinctive result is that the coefficients on height and weight for both genders are statistically insignificant in the regression.
This implies that there are useful information on body types which are embedded into various body measures. The body shapes or types cannot be fully captured by simple measures such as height or weight.

\begin{figure}[t]
     \centering
     \includegraphics[width=\linewidth]{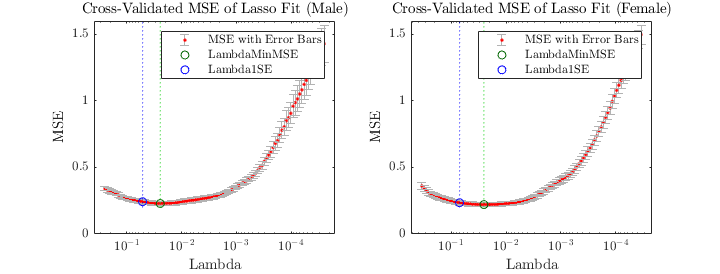}
     \vspace*{-0.3cm}
     \caption{\textbf{MSE and $\lambda$ in Lasso.}}
     \label{fig:Lasso_MSE}
 \end{figure}

Moreover, it is possible that there are interactions between different body measurements since they have close relationships to construct a body shape. So we consider the original 
regressors ($X_{i}$ and $\text{Body}_{i}$ in equation (\ref{eq:part-of-body})) and interaction terms of $\text{Body}_{i}$ as a set of regressors. This gives $797$ number of covariates, which makes the OLS regression inconsistent.\footnote{We note that OLS is consistent under some regularity conditions only if the number of observations is larger than the number of covariates.} In order to mitigate the issue of high-dimensional data, we use the following Lasso regression which is valid under a sparsity assumption:
\begin{equation}
    \min_{\psi} \left(\frac{1}{2N} \sum_{i=1}^{N}(\text{Family Income}_{i}-\psi Z_{i})^{2} + \lambda \sum_{j=1}^{p}|\psi_{j}| \right)
\end{equation}
where $Z$ is a vector of covariates including the interaction terms with size $p=797$ and $\lambda \in [0,1)$ is a regularization parameter.
We construct the lasso fit using 10-fold cross-validation. Figure \ref{fig:Lasso_MSE} plots mean-squared-error (MSE) over the sequence of the regularization parameter $\lambda$ for each gender. For males, the minimum MSE is $0.231$ at $\lambda = 0.024$ and the minimum MSE plus one standard error is $0.241$ at $\lambda = 0.047$. 
For females, the minimum MSE is $0.216$ at $\lambda = 0.017$ and the minimum MSE plus one standard error is $0.228$ at $\lambda = 0.058$. 
The regression results show that many interaction terms are statistically significant.\footnote{To save the space, we omit the results here.}  When equation (\ref{eq:part-of-body}) is re-estimated with these interaction terms (so-called "post-Lasso"), it obtains higher adjusted R squared (0.393 for males and 0.470 for females) than those in Table \ref{tb:bodymeasures-income}. Thus it is highly likely that these body measures are interrelated. 
However, constructing stylized body types based on these relevant body measures is not straightforward and a nonstandard problem.


\subsection{Physical Appearance and Graphical Autoencoder}

Characterization of geometric quantity such as physical appearance of human body shape using a sparse set of canonical features (e.g., height and weight) often causes unwanted bias and misinterpretation of data. For simple shapes like rectangles, canonical measures such as width and height already provide a complete description of the shape. Hence, shape variation among rectangles could easily be described using the two canonical parameters without much issues. However, this seldom applies to more sophisticated shape variations, if at all. Instead, the canonical shape descriptors, often hand-selected, might cause \textit{nonignorable} error in capturing genuine statistical distribution by overlooking some important geometric features or measuring highly-correlated variables redundantly, which can be thought of as a measurement error of some sort.

Unfortunately, however, extracting a complete and unbiased set of shape descriptors is not a trivial task. Furthermore, the task is highly problem-specific such that, for example, the shape descriptors for car shapes would not be appropriate for describing human body shapes. To this end, we propose a novel data-driven framework for extracting complete, unbiased shape descriptors from a set of geometric data in this paper. The proposed framework utilizes an autoencoder neural network~\citep{Bourlard1988} defined on a graphical model. In this section, we present an overview of the approach and demonstrate that the shape descriptors obtained through the new approach can actually provide a better description of data.

\subsubsection{Graphical Autoencoder}
Mathematically, human body shapes can be represented as curved surfaces, or more formally manifolds $\mathcal{M}^{(i)}$ embedded in $\mathbb{R}^3$, where $i$ is an index identifying each individual. A manifold is a topological space that locally looks like Euclidean space near each point. The statistical models that this paper concerns are, in a generic form, a regression of an economic variable $Y$ with respect to a manifold-structured regressor $\mathcal{M}$ and other covariates $X$:
\begin{equation}
    Y = \phi(\mathcal{M}, X; \theta) + \epsilon.
\end{equation}
where $\phi$ is a known function up to unknown parameter $\theta$ and $\epsilon$ is an error term. Here, a problem rises regarding the manifold regressor $\mathcal{M}$ as the regressor $\mathcal{M}$ is an abstract, geometric object and not a usual vector variable as in other typical economic and statistical models. In other words, there is no statistical model that naturally accepts the manifold regressor $\mathcal{M}$, unless $\mathcal{M}$ is somehow converted into a vector form.

Due to the above bottleneck, one may consider measuring a few geometric dimensions, such as lengths and girths, and use those measurements to encode body shapes. However, as being shown in the paper, such simplistic measurements are not an accurate characterization of complex geometric objects such as human body shapes. Instead, data driven parameterizations such as in \cite{Wang2005, Baek2012, Pishchulin2017} provide more comprehensive and reliable codification of body shapes, but many of these works assume that the human body shape distribution is linear, leading to inaccurate encoding of body shapes \citep{Freifeld:ECCV:2012, Baek2013}.

In this paper, we employ a data-driven, nonlinear parameterization of body shapes achieved via a graphical autoencoder. An autoencoder is a certain type of artificial neural network that possesses a hour-glass shaped network architecture. An autoencoder can be thought of as two multilayer perceptron (MLP) models cascaded sequentially, where the first MLP codifies a high-dimensional input into a lower dimensional embedding (encoder) and the second MLP reconstructs the original input back from the encoded embedding (decoder). Because of the dimensional bottleneck created in the middle, the neural network is promoted to search for the most effective way of compressing the high dimensional input into the lower dimensional embedding.

The concept of a graphical autoencoder we propose here is an extension of such notion of autoencoders to manifold-structured data. As in many geometric data analysis applications, we discretize a manifold $\mathcal{M}$ to a triangular mesh, achieving piece-wise linear approximation of the original surface. A triangular mesh is a graph $\mathcal{G}=\left\{ \mathcal{V}, \mathcal{E}, \mathcal{F} \right\}$ where $\mathcal{V}$ is a set of vertices/nodes, $\mathcal{E}$ are edges interconnecting the vertices, and $\mathcal{F}$ are triangular facets. We equip the meshes $\mathcal{G}_{1,\cdots,N}$ with a semantic correspondence structure such that the graph elements of the same index correspond to the same body part location across all meshes $\mathcal{G}_{1,\cdots,N}$. This process is commonly called ``registration'' or ``correspondence matching'' in computer graphics and geometry processing literature and can be achieved via methods such as \cite{zuffi2015stitched,wei2016dense,Sun2017,sun2018zernet}.

In this setting, the graphical autoencoder is defined as follows:
\begin{equation}
\begin{array}{l l}
    p = (f_1\circ f_2 \circ \cdots \circ f_m)(V\in\mathcal{V}), & \text{(encoder)} \\
    V = (g_1\circ g_2 \circ \cdots \circ g_m)(p). & \text{(decoder)}
\end{array}
\end{equation}
Here, each of the layers $f_1 \cdots f_m$ and $g_1 \cdots g_m$ are modeled as a simple perceptron:
\begin{equation}
    f_i(h) \text{ or } g_i(h) = \sigma \left( \sum_j W_i^Th + b_i \right),
\end{equation}
where $W_i$ are neural weights and $b_i$ are bias. $\sigma$ is the activation function where we empirically decide to be rectified linear unit (ReLU) activation for $f_1 \cdots f_{m-1}$ and $g_1 \cdots g_{m-1}$. We set linear activation for the terminal layers $f_m$ and $g_m$ (i.e. no rectification).

Finally, the graphical autoencoder is trained to minimize the mean square error between the original mesh and the reconstructed mesh:
\begin{equation}
    \min_{\theta_f, \theta_g} \|V-g(p)\|
\end{equation}
where $p=f(V)$ by definition, $\theta_f$ and $\theta_g$ are the model parameters of $f$ and $g$ respectively, and $V$ is the list of vertex coordinates of a graphical model.

\begin{figure}[t]
    \begin{centering}
    \includegraphics[width=\linewidth]{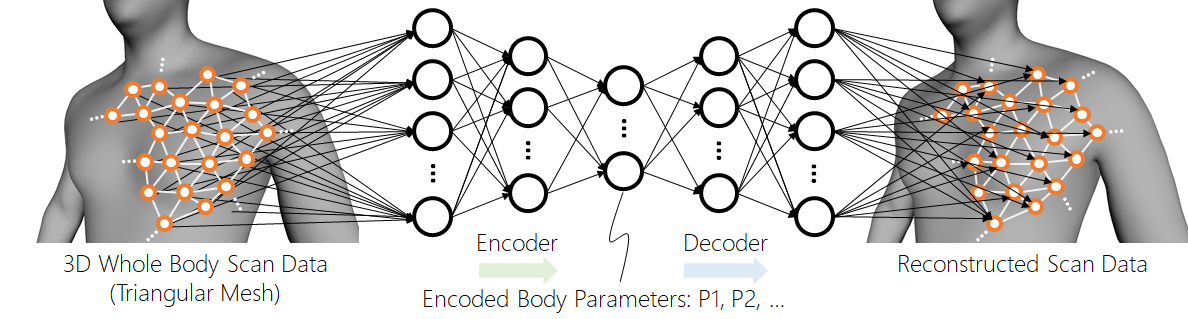}
    \end{centering}
    \vspace*{-0.3cm}
    \caption{\textbf{Schematic Illustration of the Proposed Graph Autoencoder.}
    A discrete-sampled scalar field acts as input and output nodes of the autoencoder. The intermediate layers are similar to the ordinary autoencoder layers.}
    \label{fig:graphAE}
\end{figure}

Figure~\ref{fig:graphAE} illustrates a schematic overview of the graphical autoencoder. As shown in the figure, the vertices of a topology-normalized graphical model act as input neurons in the autoencoder model. Then the input neurons are connected to the hidden neurons in the next layer which then are connected in chain through the ``bottleneck'' layer. The bottleneck layer has a significantly small number of neurons compare to the input neurons and, hence, the dimensionality compression occurs there. The latter half of the autoencoder is symmetric to the first half and finally reconstructs the bottleneck encoding into the original graphical model. The training process of the graphical autoencoder attempts to minimize the discrepancy between the reconstructed model and the original input by tuning the neural weights of the hidden layers.
 
\subsubsection{Graphical Autoencoder on CAESAR Dataset}
The CAESAR scan dataset includes $15,178$ number of vertices as well as $(x,y,z)$ coordinate. This gives us $45,534$ inputs for each individual. In order to extract body shape parameters that encode the geometric characteristics of a person's appearance, we designed a graphical autoencoder consisting of seven hidden layers. Each of the hidden layers are comprised of 256-64-16-$d$-16-64-256 neurons respectively, where $d$ is the intrinsic data dimension, or the dimensionality of the embedding. The RMSprop optimizer was used for the training. The dataset was randomly split to a training group used for training and a validation group that were set aside during the training. The ratio between the number of data samples in such groups were 80:20 respectively. The training continued until 5,000 epochs with the batch size of 200 samples. As a criterion to evaluate the performance of the graphical autoencoder, we used the reconstruction error measured in mean-squared-error (MSE). As described above, the graphical autoencoder first embeds graphical data into a lower dimensional embedding through the encoder part of the network, which then is reconstructed back into a graphical model through the decoder part. We compared how the reconstructed output is different from the original input to the network.

\begin{figure}[t]
    \begin{centering}
    \includegraphics[width=\linewidth]{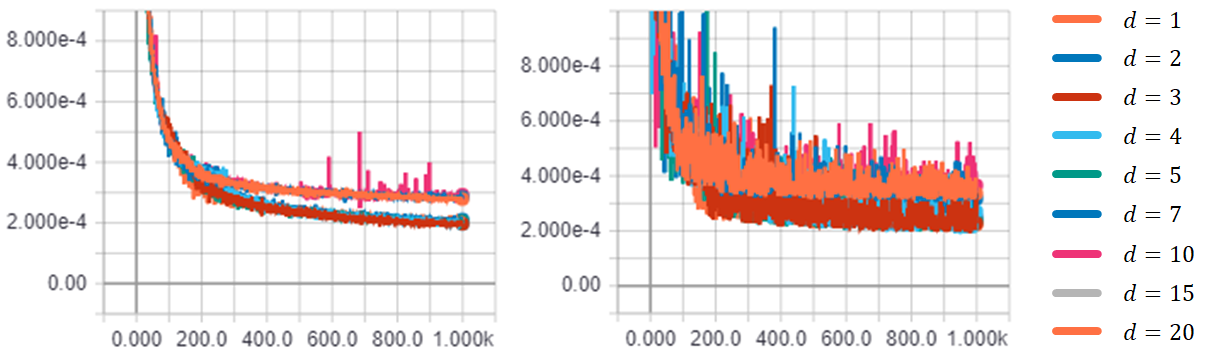}
    \end{centering}
    \vspace*{-0.3cm}
    \caption{\textbf{Result of Training Graphical Autoencoder with the Entire CAESAR Dataset.}
    The abscissa is the number of epochs for the training and the ordinate is the model loss in terms of MSE. The left shows the loss on training dataset (training loss) while the right shows the loss on validation dataset (validation loss). The accuracy did not show any significant improvement after 1,000 epochs for all cases and thus removed from the figure for the sake of better visualization.}
    \label{fig:dim_entireset}
\end{figure}

The first experiment was conducted to test the ability of the graphical autoencoder in embedding the geometric information underlying in data. To achieve this, we applied the aforementioned graphical encoder to the entire CAESAR dataset, with varying embedding dimension $d$ from 1 to 20 as reported in Figure~\ref{fig:dim_entireset}. The embedding accuracy was below $3e^{-4} \text{m}^2$ for most cases. Particularly, when the dimension $d$ was 3, it showed the lowest MSE, in both training and validation losses, which provides a justification for estimating $d=3$ as the intrinsic dimension.

For the meaning of the embedded parameters of the third dimension, the first component, $P_1$, discerned to be related to height of a person and $P_2$ to the body volume (obesity/leanness). Interestingly, as $P_3$ increases the body shape became more feminine, (namely, more prominent chest and hip-to-waist ratio) and, conversely, as it decreases the body shape became more masculine with less prominent chest and curves.

Based on such observation, we further conducted another similar experiment for training the graphical autoencoder with separate genders. Among 2,383 subjects in the CAESAR dataset, there were 1,122 males and 1,261 females. The two groups had been separated to two experiment sessions in which they were further separated to training and validation groups with the same 80:20 ratio.

\begin{figure}[t]
    \begin{centering}
    \includegraphics[width=\linewidth]{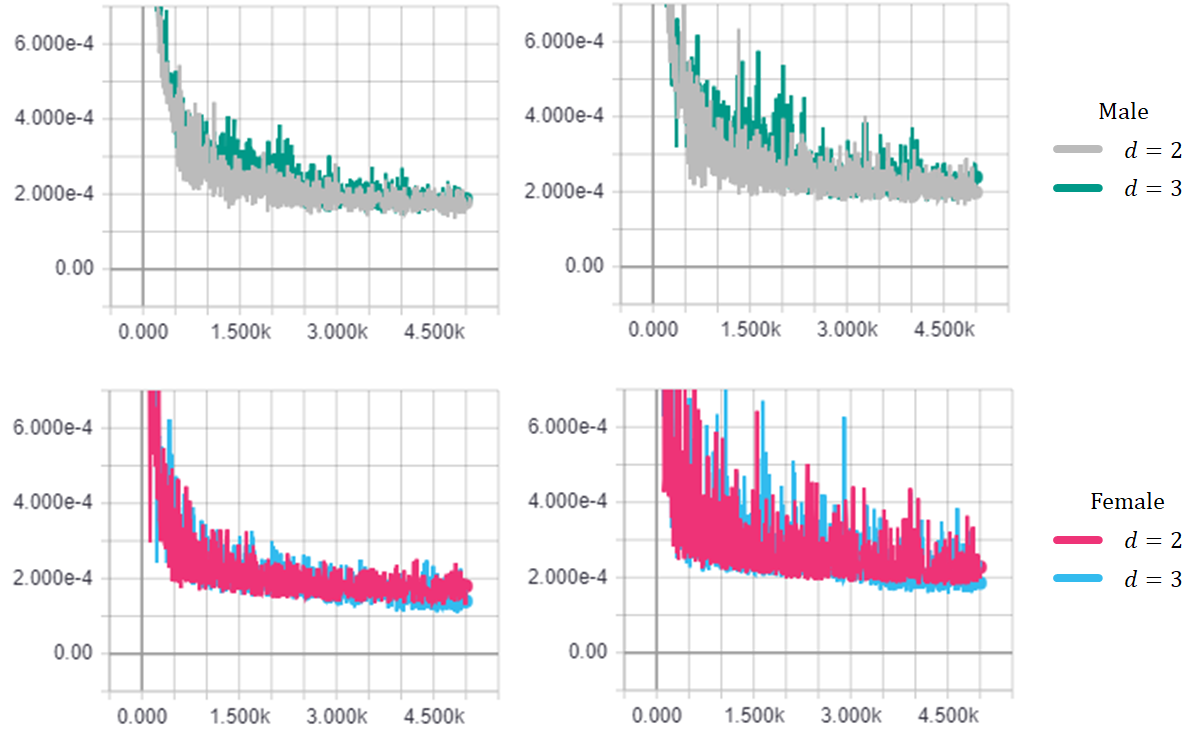}
    \end{centering}
    \vspace*{-0.3cm}
    \caption{\textbf{Result of Training Graphical Autoencoder Separately on Each Gender.}
    The abscissa is the number of epochs for the training and the ordinate is the model loss in terms of MSE. The left shows the loss on training dataset (training loss) while the right shows the loss on validation dataset (validation loss).}
    \label{fig:dim_separate}
\end{figure}

The new experiment with separate genders demonstrated a similar trend to the first experiment in terms of how the intrinsic dimension affects the reconstruction error, as visualized in Figure~\ref{fig:dim_separate}. However, interestingly, this time, the reasonable intrinsic dimension $d$ was observed to be 2 for male subjects. We interpret this result that, since now the two genders are separated, the role of $P_3$ (feminine/masculine) is now less significant than before and, thus, the gain of accuracy by including the third dimension becomes negligible for males. We also note that, however, such interpretation was not true with the female population, since the accuracy was in fact higher when $P_3$ was included. Our explanation to such is that, for the female body shapes, there is a greater variation in body curves compared to male population, and therefore, the third component has a greater significance for females. We, therefore, select $d=2$ for males and $d=3$ for females. Lastly, we also note that the convergence was slower when the two genders were separated and measurable gain of accuracy could be observed even after $1,000$ epochs, which was not the case when the two genders were combined in the training. This could be because the number of training samples in the training dataset is significantly smaller (about a half) than the previous case, rendering a drop of the representative power of the data.

\begin{figure}[t]
    \centering
    \includegraphics[width=\linewidth]{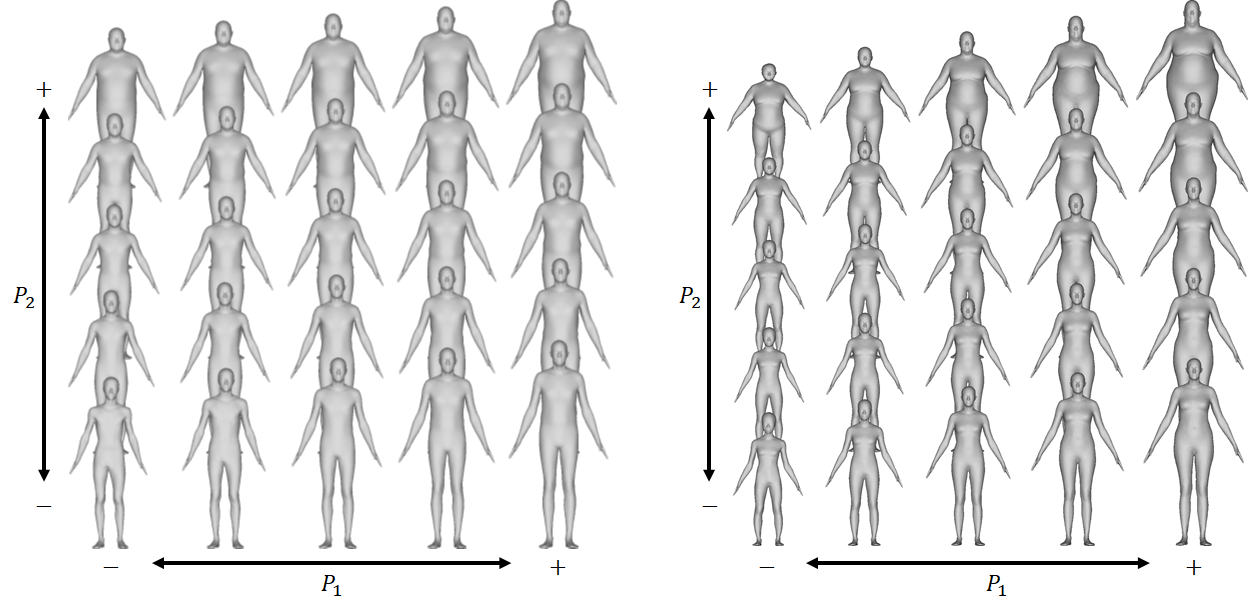}
    \vspace*{-0.3cm}
    \caption{\textbf{Body Shape Parameters Derived from the Graphical Autoencoder.}
    3D body shape models for male (left) and female (right) are arranged in accordance with their body shape parameters, with increments of -3$\sigma$, -1.5$\sigma$, 0, 1.5$\sigma$, and 3$\sigma$ with respect to the mean in each direction where $\sigma$ is the s.d. of each parameter.}
    \label{fig:shape_parm}
\end{figure}

Figure~\ref{fig:shape_parm} illustrates the body shape spanned by the two parameters obtained from the graphical autoencoder for each gender. 3D body shape models are arranged in accordance with their body shape parameters with increments of $-3\sigma$, $-1.5\sigma$, $1.5\sigma$, and $3\sigma$ with respect to the mean in each direction where $\sigma$ is the standard deviation of each parameter. Body shapes for male (left) and female (right) display similar patterns over changes in the two parameters.
Overall, the first parameter $P_1$ affects how tall a person is. That is, a smaller value in $P_1$ indicates the person is not tall compared to the other population and vice versa. $P_2$ is how lean a person is. That is, a large value in $P_2$ results in an obese person, while a small value in $P_2$ results in a more slim and fit person. 

\begin{figure}[t]
    \centering
    \includegraphics[width=\linewidth]{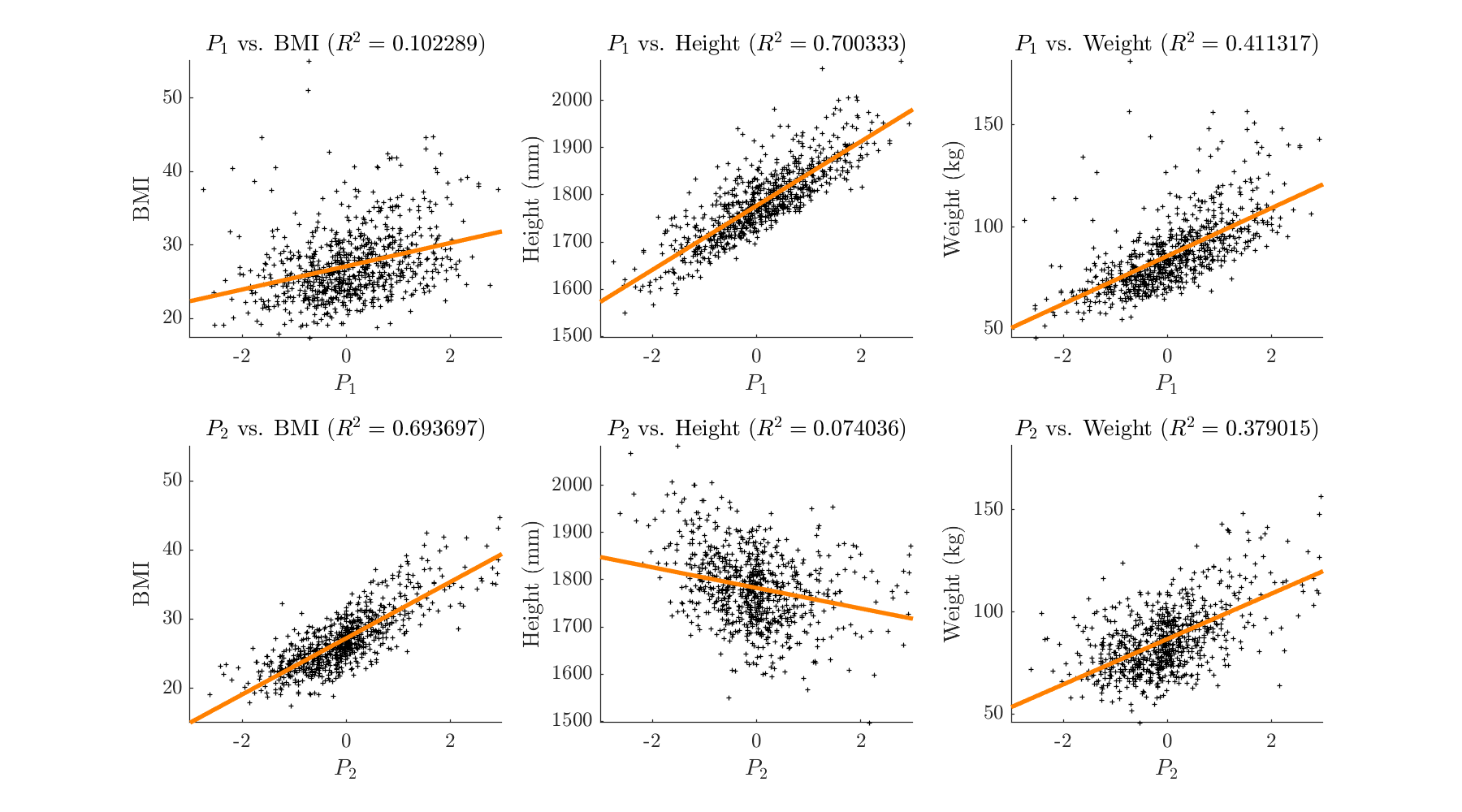}
    \vspace*{-0.3cm}
    \caption{\textbf{Relation between Body Shape Parameters and the Conventional Body Measurements for Male.}
    The straight line displays the linear fit. The R squared is reported in the parentheses. }
    \label{fig:shape_parm_male}
\end{figure}

\begin{figure}[t]
    \centering
    \includegraphics[width=\linewidth]{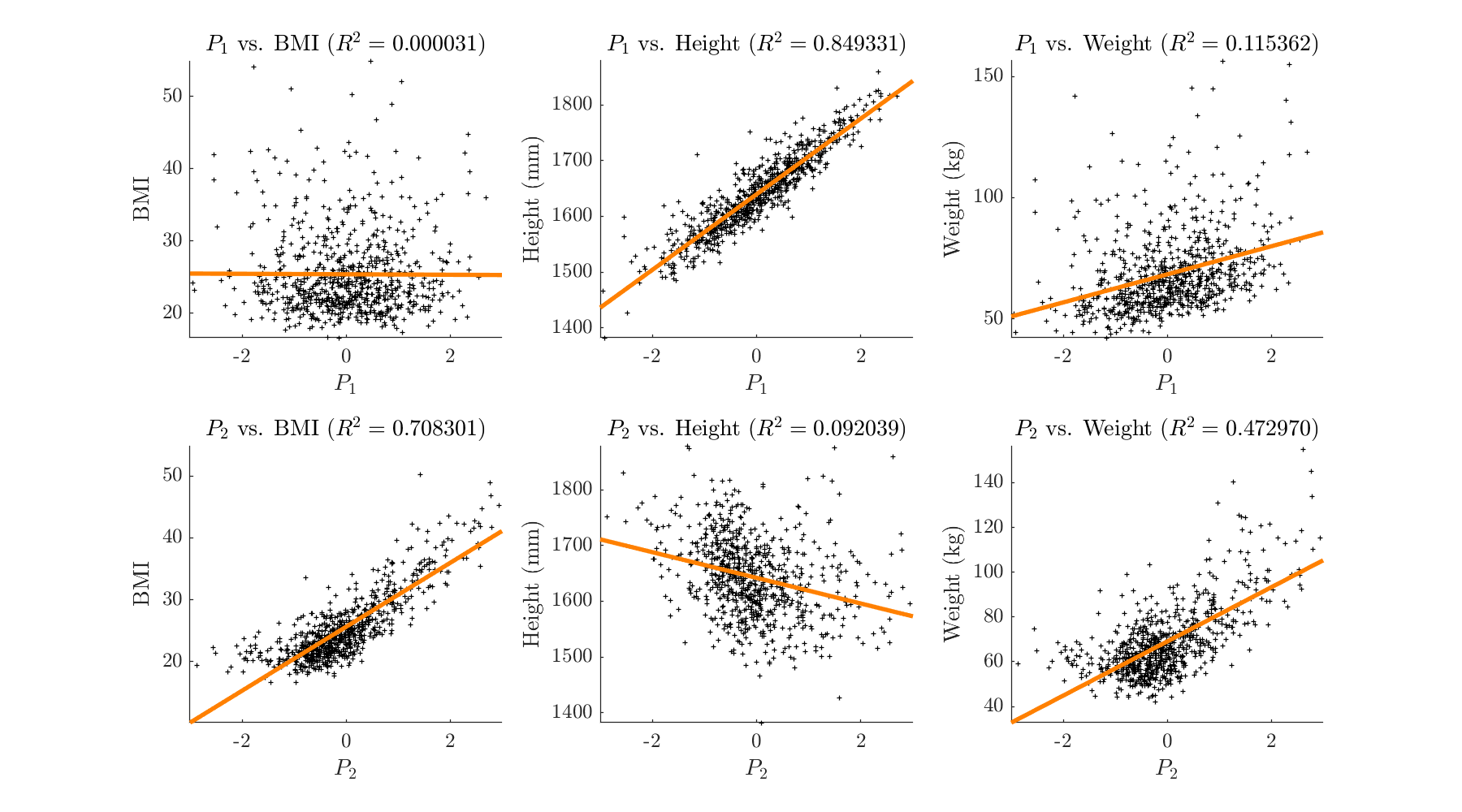}
    \vspace*{-0.3cm}
    \caption{\textbf{Relation between Body Shape Parameters and the Conventional Body Measurements for Female.}
    The straight line displays the linear fit. The R squared is reported in the parentheses. }
    \label{fig:shape_parm_female}
\end{figure}

\begin{figure}[t]
     \centering
    \includegraphics[width=0.5\linewidth]{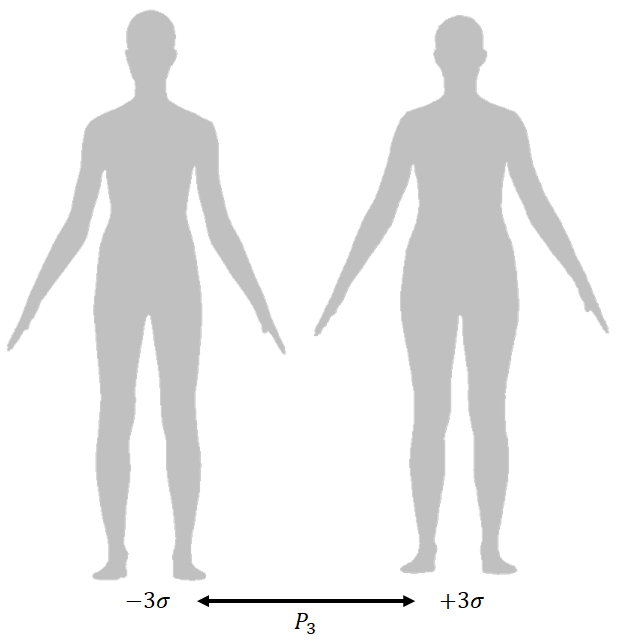}
    \vspace*{-0.3cm}
    \caption{\textbf{The Third Body Shape Parameter $P_3$ for Females.}
    The third parameter tends to capture the hip-to-waist ratio of the body shape among the female subsample.}
    \label{fig:female_p3}
\end{figure}

In order to better understand these parameters, we consider a linear fit of BMI, height, or weight on each parameter. Figure~\ref{fig:shape_parm_male} displays the relation between body shape parameters and the conventional body measurements for male. $P_1$ is positively correlated with BMI, height, and weight. Among these body measurements, height is the most highly correlated with $P_1$ (approximately $R^2 = 0.70$). $P_2$ is negatively correlated with height, but is positively correlated with BMI and weight. BMI has the highest correlation with $P_2$ (approximately $R^2 = 0.69$).
Figure~\ref{fig:shape_parm_female} displays the relation between body shape parameters and the conventional body measurements for female. The patterns are close to those for male in Figure~\ref{fig:shape_parm_male}. 
As discussed before, the female sample produces an additional feature, $P_3$. We visualize the third parameter for female  in Figure~\ref{fig:female_p3}. As shown in the figure, $P_3$ captures the ratio of  hip to waist for females, which is unique to female dataset. 
For simplicity, thus, we will interpret $P_1$, $P_2$, and $P_3$ as features associated with a person's stature, obesity, and hip-to-waist ratio, respectively.

\begin{figure}[t]
      \centering
     \includegraphics[width=0.7\linewidth]{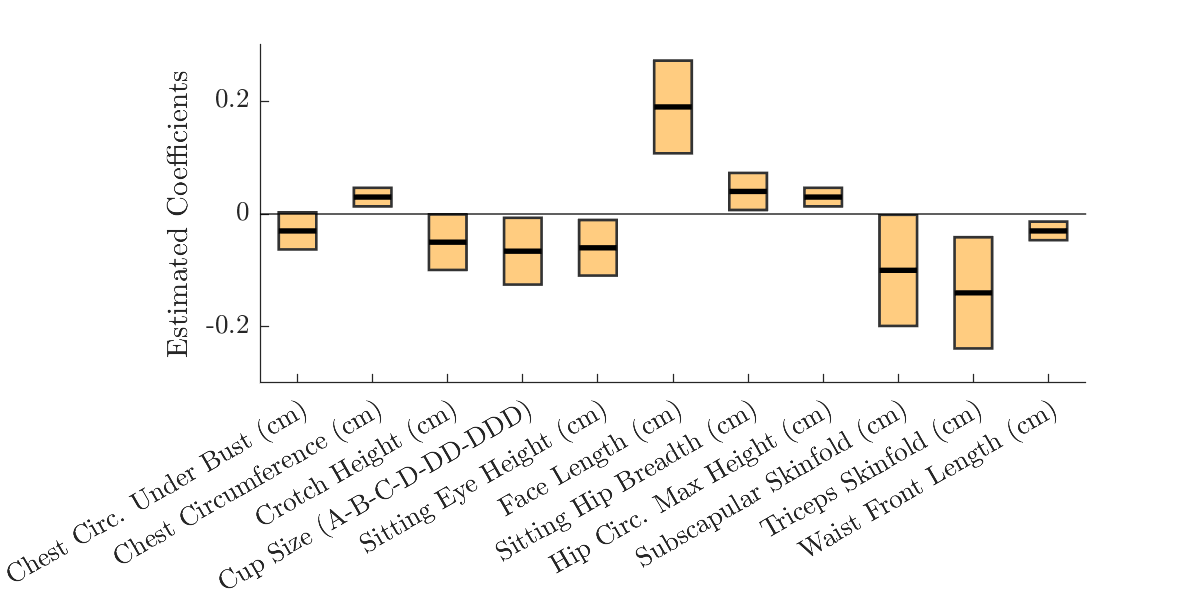}
     \caption{\textbf{Various Measures and $P_3$ for Females.}
     Estimated coefficients and bootstrapped 90\% confidence bands are reported for female. Note that units for all measurements  except cup size are converted into centimeter (cm).}
     \label{fig:various_measures_p3}
 \end{figure}

It is worthwhile to note that the extracted features $P_1$, $P_2$, and $P_3$ perform better in explaining nonlinear variations in body shapes than simple measures such as height, BMI, and hip-to-waist ratio.  As shown in Figure~\ref{fig:shape_parm}, the body shape spanned by $P_1$ and $P_2$ displays dynamic and nonlinear patterns as the parameters vary.
In addition, we consider a linear prediction of female $P_3$ using various body measures and report the results in Table \ref{tb:bodymeasures-p3} (also in Figure~\ref{fig:various_measures_p3}).\footnote{We only report statistically-significant variables.}  It shows that there are many body parts which are highly associated with  $P_3$.  This confirms that $P_3$ captures complexity in female body shapes and  reflects not only hip-to-waist ratio but also variations in other body parts. Nevertheless, we call $P_3$ hip-to-waist ratio for convenience.

\subsubsection{Extracted Body Types and Family Income}

We now use the measurements of body type which are extracted by graphical autoencoder in the previous section. We estimate the equation (\ref{eq:income-body}) with the  extracted body types in place of a set of body measurements for $\text{Body}_{i}$ as following:
\begin{align} 
\label{eq:bodyshape1}
&\text{Family Income}_{i}=\alpha X_{i} + P_{1i}  + \epsilon_{i},\\
\label{eq:bodyshape2}
&\text{Family Income}_{i}=\alpha X_{i} +  P_{2i} + \epsilon_{i},\\
\label{eq:bodyshape3}
&{\begin{cases}
\text{Family Income}_{i}=\alpha X_{i} + \beta_{1}P_{1i} + \beta_{2}P_{2i} + \epsilon_{i} & \mbox{if male},\\
\text{Family Income}_{i}=\alpha X_{i} + \beta_{1}P_{1i} + \beta_{2}P_{2i} + \beta_{3}P_{3i} +\epsilon_{i} & \mbox{if female},\\
\end{cases}}
\end{align}
where $P_{1i}$, $P_{2i}$ and $P_{3i}$ are body types for each individual $i$. Table \ref{tb:bodytype-parameters-income} reports estimation results across the gender  with the same set of controls. 
In equation (\ref{eq:bodyshape3}), we add all intrinsic features of the body shape to the income equation.  For males, only the coefficient of the $P_{1}$ measurement is statistically significant and $P_{2}$ does not explain the family income. One standard deviation increase in males' $P_{1}$ is associated with $0.052$ increase in log family income. For females, on the other hand, only the coefficient of the $P_{2}$ measurement is statistically significant, and $P_{1}$ and $P_{3}$ are not correlated with the family income. When these insignificant variables are dropped as in equations (\ref{eq:bodyshape1}) and (\ref{eq:bodyshape2}), the regression equations get higher adjusted R squared. The results show that one standard deviation decrease in females' $P_{2}$ is associated with $0.056$ increase in log family income.

For comparison, we replace the  extracted body types with height, BMI, and hip-to-waist ratio and re-estimate the above equations.\footnote{Hip-to-waist ratio is calculated as $\frac{\text{Hip Circumference, Maximum}}{\text{Waist Circumference, Preferred}} \times 100$.} The estimated results are reported in Table \ref{tb:BMI/height/hip2waist-income}. In both genders, height has positive impact on log family income and is statistically significant. The estimated coefficients of BMI and hip-to-waist ratio are insignificant. In particular, BMI was significant at 10\% significance level in Table \ref{tb:BMI-income} when height and BMI are included. However, BMI is not significant anymore when hip-to-waist ratio is added. We observe no gender differential in the impact of body types. 
The results confirm that the estimation of the income equation with conventional body measurements is susceptible to variable selection and provides different conclusions than those with our proposed method.

\subsection{Endogenous Body Types}
\subsubsection{Proxy Variables Approach}
If unobserved determinants of family income such as  individual personality and childhood nutrition are correlated with the physical appearance, the estimates in the previous section are inconsistent. To recover the causal relationship, we use the proxy variables approach where observed proxy to the unobserved determinants resolve the possible issue of endogeneity. A set of the observed proxies includes fitness, car size, birth state, and survey site.\footnote{Fitness is measured as hours of exercise per week. Car size is classified as two groups: Sedan (compact, economy, intermediate, full size, luxury, sports car) and Non-sedan (SUV, minivan, station wagon, truck, van). Birth state are classified as five groups: Foreign, West, Midwest, South, and Northeast. Survey site includes LA (CA), Detroit (MI), Ames (IA), Dayton (OH), Greensboro (NC), Marlton (NJ), Ottawa (Ontario, CAN), Minneapolis (MN), Houston (TX), Portland (OR), San Francisco (CA), and Atlanta (GA).} 
We choose these variables as relevant proxies since fitness and car size could be related to individual preference to body types, and birth state and survey site could reflect local nutrition environments.
To control for the endogeneity, thus, we assume the conditional independence of body types and unobserved determinants of family income, conditional on the observed proxies. 

We estimate equation (\ref{eq:bodyshape3}) by controlling for  various subset of the proxy variables. 
As a comparison, we use measured Height, BMI, and hip-to-waist Ratio in place of $P_1$, $P_2$, and $P_3$, respectively. The estimation results are reported in Table  \ref{tb:BMI/Height/Hip-to-waist-Ratio-income-endogeneity-PV-approach}. The estimated coefficients of height are similar to those in columns for equation (\ref{eq:BMI3}) for both genders in Table \ref{tb:BMI-income}. In particular, height still has positive effects on family income in both genders. 
On the other hand, the estimated coefficient of BMI for female becomes significant; BMI has negative effects in female subsample. 

Table \ref{tb:bodytype-parameters-income-endogeneity-PV-approach}  reports estimation results  for the equation   (\ref{eq:bodyshape3}) with $P_1$, $P_2$, and $P_3$. It is worth noting that fitness and car size are not statistically significant but birth state (Northeast) and survey site (Dayton, OH for male; Marlton, NJ for female) are statistically significant. When all proxy variables are included, the estimated coefficient of $P_1$ is $0.056$ for males. It is statistically significant   at 1\% significance level. 
Thus taller males have a tendency to have higher family income. But we do not find statistically meaningful relationship between the males' obesity and the family income.
For females, the $P_{2}$ measurement is negatively associated with the family income  and its coefficient is statistically significant at 1\% significance level. Thus we find that females' obesity  matters for the family income in a negative sense but their stature and hip-to-waist ratio are not associated with the family income. The results are qualitatively similar to those from
Table \ref{tb:bodytype-parameters-income}.

\subsubsection{Control Functions Approach}

Even though some unobserved determinants of family income  have been controlled for, there would be other possible unobserved factors such as individual ability. As reported in Persico et al. (2004), Case and Paxson (2008), and Lundborg et al. (2014), stature would be highly correlated with individual cognitive and noncognitive abilities. 
Thus we conjecture $P_1$ or stature would be possibly endogenous regressor. To address this issue, we look for a set of valid instrumental variables (IV). Our identification strategy is to assume that 
unobserved determinants of Shoe size, Pants size, and Jacket Size for male (or Blouse Size for female) are uncorrelated with individual cognitive and noncognitive abilities. Given the condition, we adopt these unobserved size determinants or variations as valid IVs and incorporate the instrumental variables or control functions (CF) approach into the economic model based on the graph convolution method. 

\begin{figure}[t]
    \centering
    \includegraphics[width=0.6\linewidth]{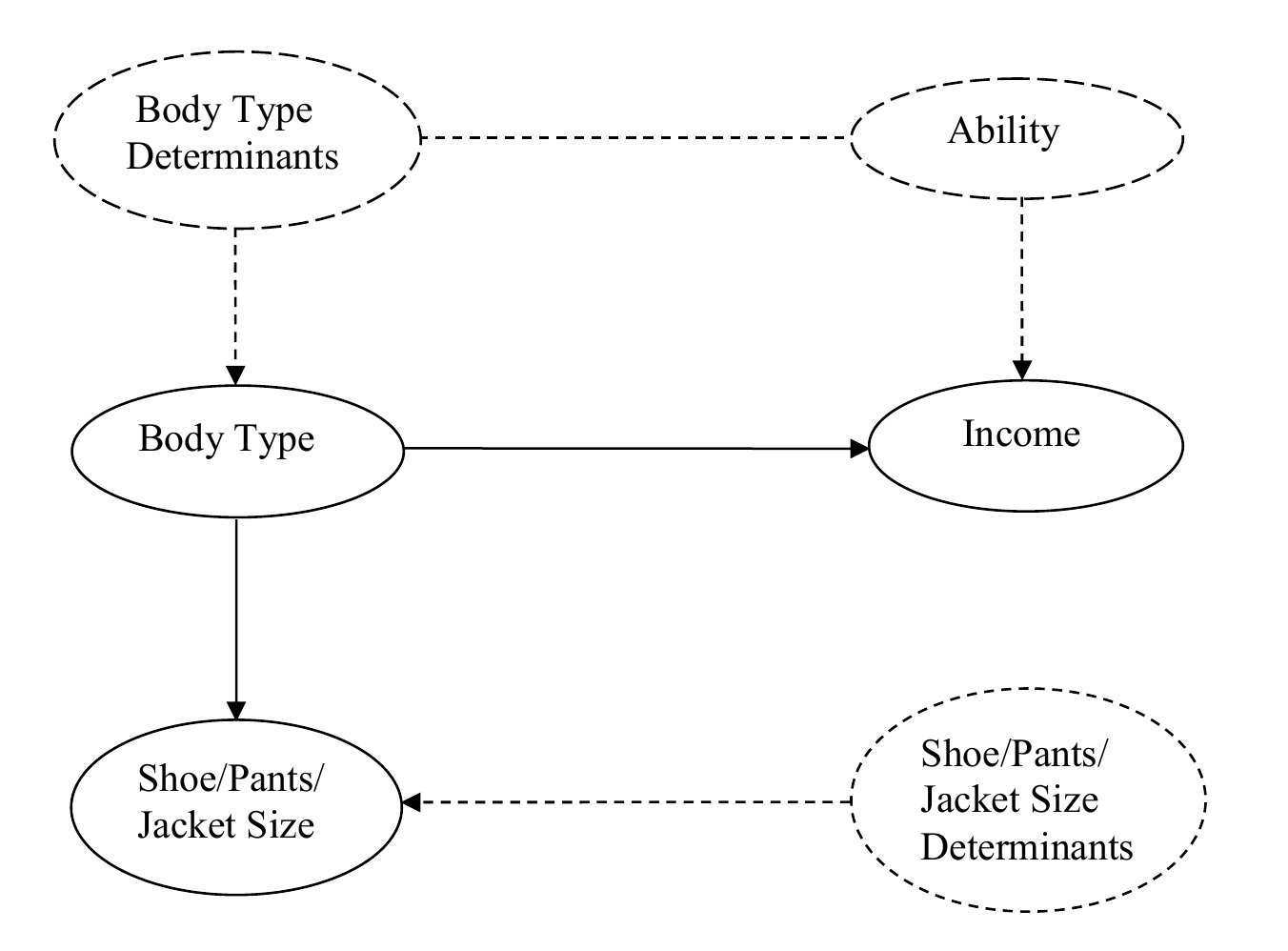}
    \vspace*{-0.3cm}
    \caption{\textbf{Causal Diagram for the Impact of Body Type on Income.}
    Shoe/Pants/Jacket size determinants are independent of individual ability so that they can serve as valid instrumental variables.}
    \label{fig:causal_diagram}
\end{figure}

Figure~\ref{fig:causal_diagram} shows a graphical depiction of the causal diagram (see e.g. \citet{Pearl00} and \citet{ChalakWhite11}). Complete circles denote observed variables and dashed circles denote unobserved determinants. Arrows denote direct causal relations. A line without arrows denotes dependence between two variables. The primary parameter of interest is the impact of body type on family income. But body type (stature) is endogenous due to the dependence of body type determinants and cognitive and noncognitive abilities. Shoe size, jacket (blouse) size, and pants size are determined by body type as well as unobserved determinants such as personal size preference. We assume that such unobserved size determinants are uncorrelated with ability. Given the condition, they can serve as legitimate IVs to identify the causal relation. Indeed, it is plausible to assume that these variables are legitimate IVs. First, they are unlikely to be correlated with the unobserved determinants of family income such as cognitive and noncognitive abilities. For instance, it is natural to assume that highly able people do not necessarily wear bigger shoes or pants given their foot length or waist circumference. In fact, it is unlikely that these IVs directly cause the family income. Second, as shown in the estimation results below, they are also strongly correlated with the body type. 

Since each size determinants or variations are unobserved, we estimate them from the projection of the observed size on the most relevant body part. In practice, we consider the projection of the reported shoe size, pants size and jacket (or blouse size) on the measured foot length, waist circumference, and chest circumference, respectively. The residuals from each projection are used as the estimated size determinants or variations (analogous to the `residuals as instruments' in \citet{HausmanTaylor83}).

We consider the following first-step reduced form equation:
\begin{align} 
\label{eq:cf-1st-stage-equation1}
P_{1i}=\delta X_{i} &+ \gamma_{1}\text{Shoe Size Determinants}_{i} \nonumber\\
&+ \gamma_{2}\text{Jacket Size/Blouse Size Determinants}_{i} \\
&+ \gamma_{3}\text{Pants Size Determinants}_{i} + \nu_{i}, \nonumber
\end{align}
where $\text{Shoe Size Determinants}_{i}$ is the estimated individual $i$'s variation or determinants in shoe size and $\text{Jacket Size/Blouse Size Determinants}_{i}$ is the estimated variation in jacket size for male (or blouse size for female), and $\text{Pants Size Determinants}_{i}$ is variation in pants size, and where $X_{i}$ are a set of exogenous regressors and $\nu_{i}$ is idiosyncratic shocks to the $P_{1i}$.
By construction, $\nu_{i}$ is the component which generates the endogeneity. From the reduced-from equations (\ref{eq:cf-1st-stage-equation1}), we estimate the control function $\hat{\nu}_{i}$.
In the second-step, we then estimate the income equation by adding the control function as following:
\begin{align} 
\label{eq:cf-2nd-stage-equation}
&{\begin{cases}
\text{Family Income}_{i}=\alpha X_{i} + \beta_{1}P_{1i} + \beta_{2}P_{2i} + \pi\hat{\nu}_{i}  + \epsilon_{i} & \mbox{if male},\\
\text{Family Income}_{i}=\alpha X_{i} + \beta_{1}P_{1i} + \beta_{2}P_{2i} + \beta_{3}P_{3i} + \pi\hat{\nu}_{i}  + \epsilon_{i} & \mbox{if female}.
\end{cases}}
\end{align}
Since $\hat{\nu}$ corrects for the sources of the endogeneity, we can consistently estimate the parameters associated with the physical appearances. Other nice feature of the control functions approach is that we can test whether the physical appearances are endogenous by checking if $\pi=0$.

Table \ref{tb:bodytype-parameters-income-endogeneity-CF-approach}  reports estimation results  for the equations (\ref{eq:cf-1st-stage-equation1})-(\ref{eq:cf-2nd-stage-equation}).
In the columns for the equation (\ref{eq:cf-1st-stage-equation1}), all IVs are statistically significant and positively correlated with stature in both genders. Experience is also positively associated with individual's stature and the relation is nonlinear. 
People who were born in Foreign countries or Northeast are likely to be shorter than those born in Midwest.
For males, education is negatively correlated with stature. Asian people are less likely to be taller than White people. For females, Hispanic and Asian people are less likely to be taller than White people.

In the columns for the equation (\ref{eq:cf-2nd-stage-equation}),
the estimated coefficient of $P_1$ is $0.097$ for male, which is larger than that from Table \ref{tb:bodytype-parameters-income-endogeneity-PV-approach}. It is statistically significant at 5\% significance level. 
Thus taller males have a tendency to have higher family income. Interestingly, we do not find statistically significant causal relationship between males' obesity and the family income.
We estimate that one standard deviation increase in $P_{1}$ measurement is associated with $\$0.097\times 70,000 = \$6,790$ increase in the family income for a male who earns $\$70,000$ of median family income. This is equivalent to $\$\frac{0.097}{6.8} \times 70,000=\$998.5$ increase of family income per centimeter. 
Note that for males one standard deviation in $P_{1}$ is equivalent to $6.8$ centimeter in height and one standard deviation in $P_2$ is equivalent to $4.07 kg/m^2$ in BMI.

The estimation results for the covariates resemble those in previous tables. As shown in the literature on the returns to education, education has a positive impact on family income. Its estimated coefficient is $0.046$ and statistically significant at 1\% significance level. Males born in Northeast have tendency to have higher family income than those born in Midwest.
Interestingly, the estimated coefficient of $\hat{\nu}_{1}$ is statistically insignificant. Thus, we find no strong evidence that $P_{1}$ is a endogenous regressor. 

For females, the estimated coefficient of $P_{1}$ is negative and the estimated coefficient of $P_{3}$ is positive, but they are not statistically significant. The $P_{2}$ measurement is negatively associated with the family income.  Its coefficient is $-0.069$ and statistically significant at 1\% significance level. Thus we find that a female's obesity negatively matters for her family income, but we do not find a strong causal impacts of her stature and hip-to-waist ratio on the family income.
One standard deviation decrease in $P_{2}$ measurement is associated with $\$0.069 \times 70,000 = \$4,830$ increase in the family income for a female who earns $\$70,000$ family income. This can be interpreted to $\$\frac{0.069}{5.17} \times 70,000 = \$934.2$ increase per one unit of BMI. Note that for females one standard deviation in $P_{1}$ is equivalent to $6.8$ centimeter in height and one standard deviation in $P_2$ is equivalent to $5.17 kg/m^2$ in BMI.  

For females, experience is important to have higher family income. As commonly reported in the literature on the wage equation, the experience displays a quadratic functional form.
Education has a positive impact on the family income and its coefficient is statistically significant at 1\% significance level, which is similar to the finding in the male case. Similarly, females born in Northeast have tendency to have higher family income than those born in Midwest. The estimated coefficient of $\hat{\nu}$ is statistically significant. Thus, we find substantial evidence that stature is endogenous in the female's income equation.

As a comparison, we apply the control functions approach to the income equations where height, BMI, and hip-to-waist ratio are used in place of the extracted body features. The corresponding reduced-form and structural equations are the same as the equations (\ref{eq:cf-1st-stage-equation1})-(\ref{eq:cf-2nd-stage-equation}).
Table \ref{tb:BMI/Height/Hip-to-waist-Ratio-income-endogeneity-CF-approach}  reports estimation results. For male, we estimate that one centimeter increase in height is associated with $\$0.01 \times 70,000 = \$700$ increase in the family income for a male who earns $\$70,000$ of median family income. The estimated effect is smaller than that when the extracted features from the deep learning are used. 
For females, pants size is not associated with height in the first step regression. The estimated coefficient of $\hat{\nu}$ 
is significant in female sample, which implies evidence of endogenous female height.  For females, interestingly, coefficients of height, BMI, and hip-to-waist ratio are all insignificant. Thus we do not find strong evidence of body shape effects when height, BMI, and hip-to-waist ratio are used.
As a result, we observe that the estimation results with the conventional measurements are volatile across different regression models -- OLS, proxy variable approach, and control functions approach. On the other hands, those with the deep-learned body parameters are very stable across different models and interestingly captures gender differential in the impact of body types on income.

\begin{figure}[t]
     \centering
     \includegraphics[width=\linewidth]{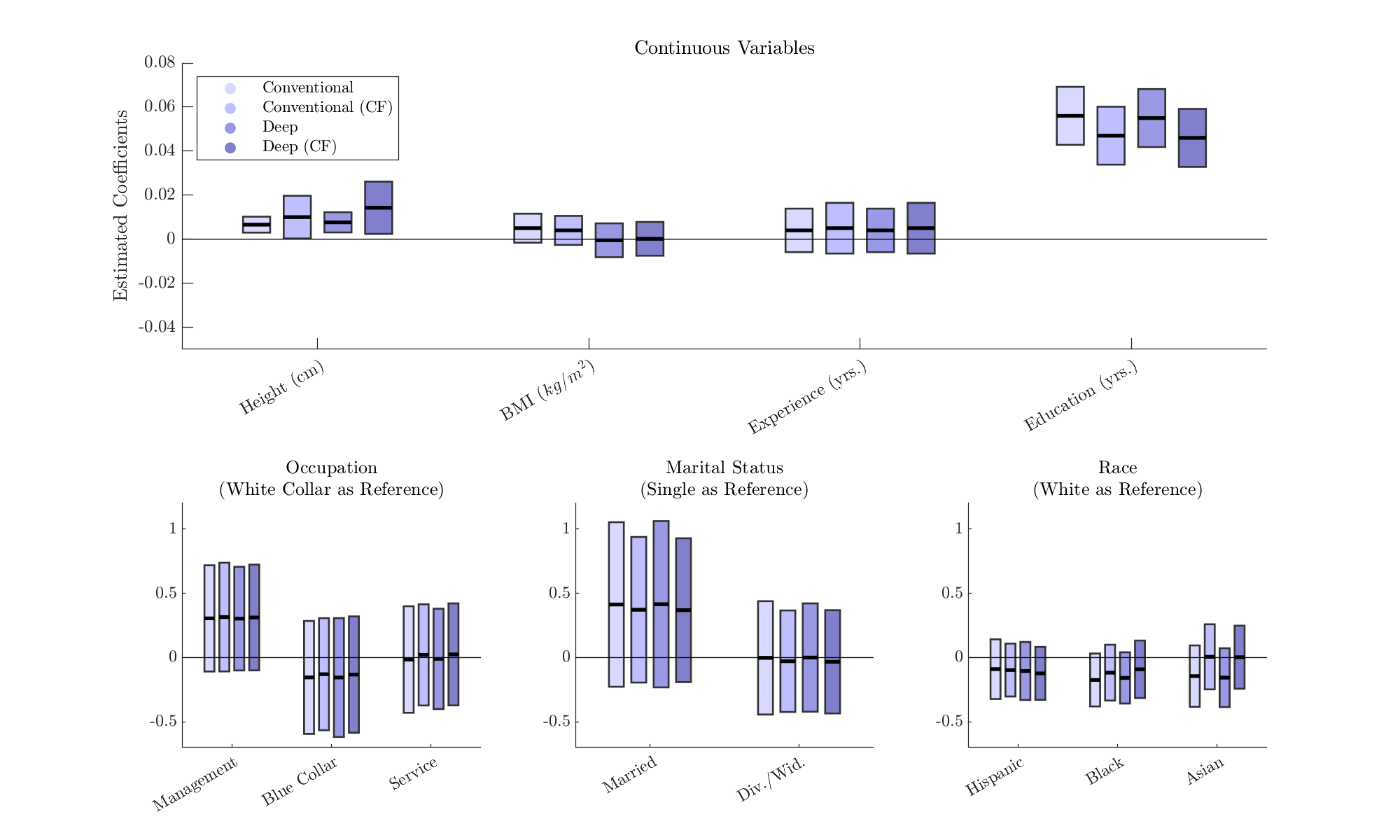}
     \vspace*{-0.3cm}
     \caption{\textbf{Comparison of Conventional Body Measures and Deep-learned Body Parameters (Male).}
     Estimated coefficients and bootstrapped 90\% confidence bands are reported. We provide results from conventional body measures (Conventional) and deep-learned body parameters (Deep) with/without control function (CF) approach. Note that the unit for height is converted into centimeter (cm).}
     \label{fig:deep_vs_conv_male}
 \end{figure}

\begin{figure}[t]
     \centering
     \includegraphics[width=\linewidth]{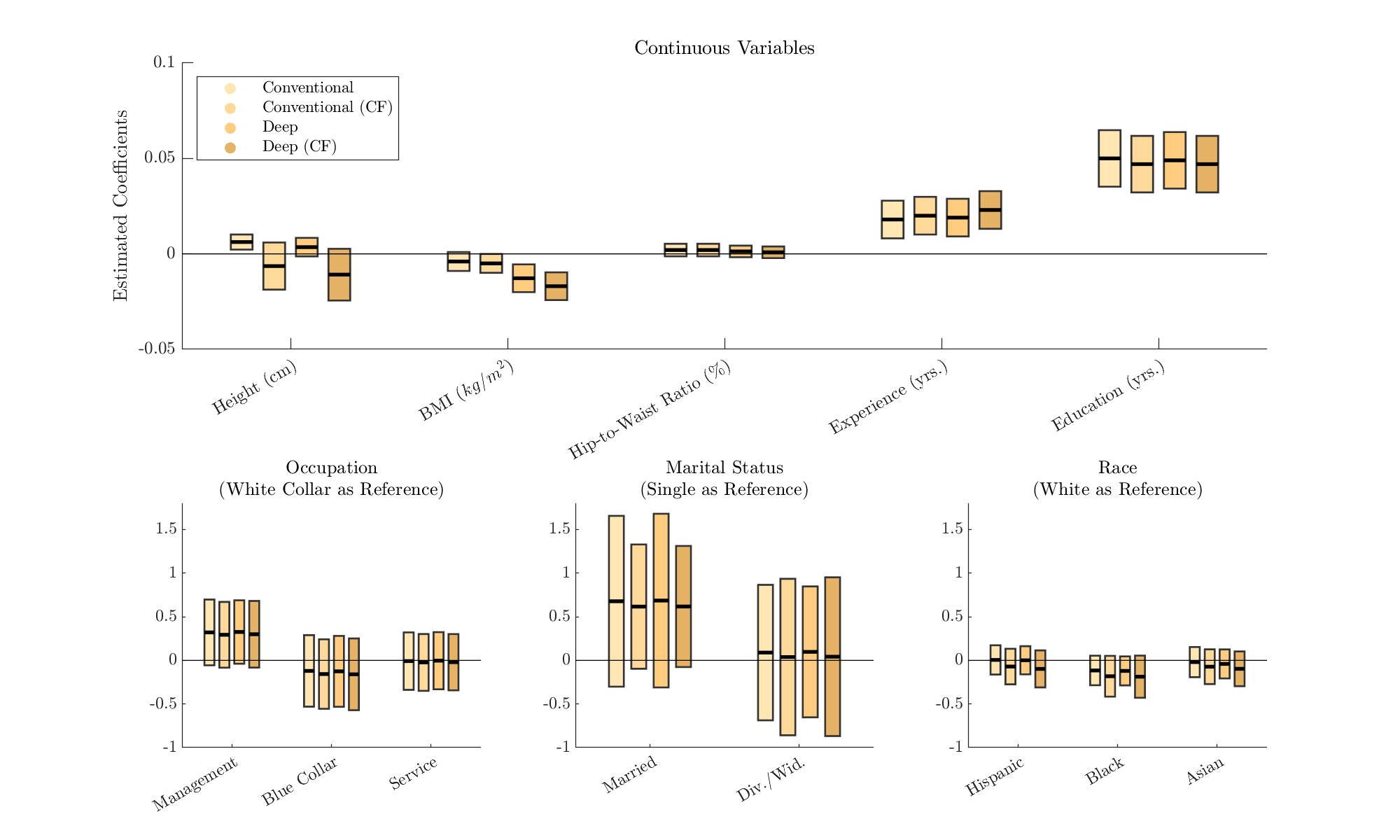}
     \vspace*{-0.3cm}
     \caption{\textbf{Comparison of Conventional Body Measures and Deep-learned Body Parameters (Female).}
     Estimated coefficients and bootstrapped 90\% confidence bands are reported. We provide results from conventional body measures (Conventional) and deep-learned body parameters (Deep) with/without control function (CF) approach. Note that the unit for height is converted into centimeter (cm).}
     \label{fig:deep_vs_conv_female}
 \end{figure}

Finally, we summarize the estimated coefficients in Figure \ref{fig:deep_vs_conv_male} for males and in Figure \ref{fig:deep_vs_conv_female} for females.  We compare the results from the conventional body measures to those from the deep-learned body parameters with/without controlling for the endogeneity.  
They show that the estimated effects from the measured height, BMI, and hip-to-waist ratio are substantially different than those from the deep learned parameters. One can possibly interpret such difference as a limitation of conventional body measures on describing appearances. In fact, it is widely known in literature (CDC document, accessed 2019\nocite{cdcbmi}) that BMI is a surrogate measure of body fatness. Neither does it distinguish fat, muscle, or bone mass, nor does it describe distribution of fat among people. As such, there is a chance where the difference in family income is falsely correlated to stature while the true underlying statistics suggests otherwise. To illustrate this problem, consider a tall and visually obese male and a short and muscular male with the same body mass. In this case, since there is no difference in BMI, the difference in family income must be explained by stature, which may lead to an inaccurate conclusion. This, however, was not the case for the deep-learned parameters.

\section{Conclusion}

This paper studies the relationship between the physical appearance and family income. We show there are significant reporting errors in the reported height and weight, and show that these discrete measurements are too sparse to provide complete description of the body shape. In fact, these reporting errors are shown to be correlated with individual backgrounds. We also find that the regression of family income on the self-reported measurements suffers from the issue of reporting errors and delivers biased estimates compared to the regression on the true measurements. The findings shed light on the importance of measuring body types instead of simply relying on subjects' self-reports for public policies. 

We introduce a new methodology built on graphical autoencoder in deep machine learning. From the three dimensional whole-body scan data, we identify two intrinsic features consisting of human body shapes for males and three intrinsic features for females. These body features are presumably less likely to suffer from measurement errors on the physical appearances. We also take into account  a possible issue of endogenous body shapes by utilizing proxy variables and control functions approaches. The empirical results document positive impact of stature and negative impact of obesity on family income for males. On the other hand, results for females show that obesity is the only significant feature and it negatively affects family income. The findings support the hypotheses on the physical attractiveness premium and the differential treatment across the gender in the labor market outcomes. 
Finally, we believe that the proposed method can be applied to many interesting research questions in Economics which deal with geometric data such as graphical, image, spatial, and social networks data.

\bibliography{shape_matters}

\newpage
\appendix 

\section{Appendix --- Empirical Results}
\begin{table}[H]
\centering
\begin{tabular}{ l c c c c c }
        \hline
        \hline\\
        Variable & Mean & Median & S.D. & Min & Max \\[2ex]
        \hline\\
Family Income (\$) & 76,085 &70,000 & 41,470 & 7,500 & 150,000 \\ [2ex]

Reported Height (mm) &1,798.2  & 1,803.4 & 82.5 & 1,498.6 & 2,108.2 \\ [2ex]

Reported Weight (kg) & 86.0 & 83.9 & 17.3 & 48.5 & 188.2 \\ [2ex]

Reported BMI (kg/m$^2$) &  26.5 & 25.8 & 4.6 & 14.0 &  59.5\\ [2ex]

Height (mm) & 1,782.6 & 1,778.5 & 78.1 & 1,497.0 & 2,084.0 \\ [2ex]

Weight (kg) &  86.8 & 83.9 & 17.5 & 45.8 & 181.4 \\ [2ex]

BMI (kg/m$^2$) &  27.2 &26.4 & 4.8 & 17.4 &  55.1\\ [2ex]

Experience (years) &   17.5 & 17.0 & 10.2 & 0 &  47.0\\ [2ex]

Education (years) &16.3 &  16.0 &  2.5& 12.0 & 24.0 \\ [2ex]

\# of Children &1.3 & 1.0 & 1.4 & 0 & 7.0\\ [2ex]

Fitness (hours) & 4.2& 2.5   & 3.0 & 0.5 &  10.0\\ [2ex]

\hline\\
       Variable  &  \# of Samples &  Variable &  \# of Samples&  &  \\[2ex]
         \hline\\

Marital Status (Single) & 240 & Race (White) & 644 & & \\ [2ex]

Marital Status (Married) & 473 & Race (Hispanic) & 18 & & \\ [2ex]

Marital Status (Div./Wid.) & 61 & Race (Black)& 68& & \\ [2ex]

Occupation (White Collar) & 461 & Race (Asian)& 44 & & \\ [2ex]

Occupation (Management) & 144 & & & & \\ [2ex]

Occupation (Blue Collar) & 101 & & & & \\ [2ex]

Occupation (Service) & 68 & & & & \\ [2ex]





Birth Region (Foreign) &  159 &  & & &  \\ [2ex]

Birth Region (Midwest) &  275 &  & & &  \\ [2ex]

Birth Region (Northeast) &  106&  & & &  \\ [2ex]

Birth Region (South) & 106 &  & & &  \\ [2ex]

Birth Region (West) & 128 &  & & &  \\ [2ex]

\hline \\
\# of Total Observations & 774 & & & &   \\ [2ex]
 \hline 
        
    \end{tabular}
    \vspace{0.3cm}
    \caption{Summary Statistics (Male)}
    \label{tb:summarystat-men}
\end{table}

\begin{table}[H]
\centering
    \begin{tabular}{ l c c c c c }
        \hline
        \hline\\
        Variable & Mean & Median & S.D. & Min & Max \\[2ex]
        \hline\\
Family Income ( \$) & 65,998 & 52,500  & 38,853 & 7,500 & 150,000 \\ [2ex]

Reported Height (mm) & 1,649.6 & 1651.0 & 76.1 & 1,320.8 & 1,930.4\\ [2ex]

Reported Weight (kg) & 67.9 & 63.5 & 16.9 & 37.2 & 172.3  \\ [2ex]

Reported BMI (kg/m$^2$) & 24.9 & 23.3 &   5.9 & 12.9 &  57.8\\ [2ex]

Height (mm) & 1,642.2 & 1,640.0 & 71.3 & 1,382.0 & 1,879.0  \\ [2ex]

Weight (kg) & 68.8 &64.9 & 17.3 & 39.2 & 156.5 \\ [2ex]

BMI (kg/m$^2$) & 25.5  &23.8  & 6.1 & 15.2 & 57.1 \\ [2ex]

Experience (years) & 18.6 & 19.0 & 10.8 & 0& 50.0 \\ [2ex]

Education (years) &  15.8 & 16.0 & 2.1 & 12.0 & 24.0 \\ [2ex]

\# of Children & 1.0 & 0 & 1.2 & 0& 6.0 \\ [2ex]

Fitness (hours) & 3.7 & 2.5 & 2.7 & 0.5 & 10.0 \\ [2ex]

\hline\\
       Variable  &  \# of Samples & Variable & \# of Samples &  &  \\[2ex]
        \hline\\

Marital Status (Single) & 248 & Race (White) & 644 & & \\ [2ex]

Marital Status (Married) & 407 & Race (Hispanic) & 11 & & \\ [2ex]

Marital Status (Div./Wid.) & 134 & Race (Black) & 88 & & \\ [2ex]

Occupation (White Collar) & 607 & Race (Asian)& 46& & \\ [2ex]

Occupation (Management) & 52 & & & & \\ [2ex]

Occupation (Blue Collar) & 49 & & & & \\ [2ex]

Occupation (Service) & 81 & & & & \\ [2ex]





Birth Region (Foreign) &  105  &  & & &  \\ [2ex]

Birth Region (Midwest) &  318  &  & & &  \\ [2ex]

Birth Region (Northeast) &  103 &  & & &  \\ [2ex]

Birth Region (South) & 122 &  & & &  \\ [2ex]

Birth Region (West) & 141 &  & & &  \\ [2ex]

\hline \\
\# of Total Observations & 789 & & & &   \\ [2ex]
 \hline 

    \end{tabular}
    \vspace{0.3cm}
    \caption{Summary Statistics (Female)}
    \label{tb:summarystat-women}
\end{table}

\begin{table}[H]
\centering
    \begin{tabular}{ c| c  }
        \hline
        \hline\\
        Variable (mm)& Variable (mm)\\[2ex]
        \hline\\
Acromial Height, Sitting &  Head Length  \\ [2ex]
Ankle Circumference  &  Hip Breadth, Sitting \\ [2ex]
\makecell{Arm Length \\(Spine to Wrist)}  &  Hip Circumference, Maximum \\ [2ex]
\makecell{Arm Length \\(Shoulder to Wrist)} &   Hip Circumference Max Height\\ [2ex]
\makecell{Arm Length \\(Shoulder to Elbow)}  &   Knee Height\\ [2ex]
\makecell{Armscye Circumference \\(Scye Circumference Over Acromion)}  &  Neck Base Circumference \\ [2ex]
Bizygomatic Breadth  &  Shoulder Breadth \\ [2ex]
Chest Circumference  &   Sitting Height\\ [2ex]
Bust/Chest Circumference Under Bust  &   Height\\ [2ex]
Buttock-Knee Length  &  Subscapular Skinfold \\ [2ex]
\makecell{Chest Girth at Scye \\(Chest Circumference at Scye)} &  Thigh Circumference \\ [2ex]
Crotch Height  &   Thigh Circumference Max Sitting\\ [2ex]
Elbow Height, Sitting  &  Thumb Tip Reach \\ [2ex]
Eye Height, Sitting  &  Triceps Skinfold  \\ [2ex]
Face Length  &   \makecell{Total Crotch Length \\(Crotch Length)}\\ [2ex]
Foot Length  &  Vertical Trunk Circumference\\ [2ex]
Hand Circumference &   Waist Circumference, Preferred\\ [2ex]
Hand Length  &   Waist Front Length\\ [2ex]
Head Breadth  &   Waist Height, Preferred\\ [2ex]
Head Circumference  &    Weight (kg)\\ [2ex]
\hline
    \end{tabular}
    \vspace{0.3cm}
    \caption{List of Various Body Measures}
    \label{tb:list-of-bodymeasures}
\end{table}

\begin{table}[H]
\centering
\begin{tabular}{c|c|c|c|c }
\hline
\hline
\multirow{3}{*}{Variable} & \multicolumn{2}{c|}{Error in Height (Eq. (\ref{eq:ReportingError-Height}))} & 
    \multicolumn{2}{c}{Error in Weight (Eq. (\ref{eq:ReportingError-Weight}))} \\
\cline{2-5}
 & Male & female & Male & female \\
\hline
Intercept & \makecell{114.481*** \\ (36.056)} & \makecell{41.826 \\ (38.067)} 
& \makecell{8.574**\\(3.438)} 
& \makecell{4.056*\\ (2.329)} \\[2ex]

\makecell{Height \\ (mm)} & \makecell{-0.006 \\ (0.011)} & \makecell{0.004 \\ (0.019)} 
& \makecell{ \\} 
& \makecell{\\ } \\ [2ex]

\makecell{Weight \\ (kg)} & \makecell{ \\ } & \makecell{ \\ } 
& \makecell{-0.056*** \\(0.018)} 
& \makecell{-0.040***\\ (0.015)} \\ [2ex]

\makecell{Family Income \\} & \makecell{-5.847*** \\ (2.738)} & \makecell{-0.105 \\ (1.810)} 
& \makecell{-0.477 \\(0.365)} 
& \makecell{-0.112\\ (0.194)} \\ [2ex]

Age & \makecell{-1.067 \\ (0.728)} & \makecell{-1.801* \\ (0.920)} 
& \makecell{0.074 \\(0.111)} 
& \makecell{0.018\\ (0.061)} \\ [2ex]

Age$^2$ & \makecell{0.014* \\ (0.008)} & \makecell{0.021** \\ (0.011)} 
& \makecell{-7.6e-4\\(0.001)} 
& \makecell{-1.0e-6\\ (7.1e-4)}\\
\end{tabular}
    \vspace{0.3cm}
    \caption{The Association between Reporting Error in Height/Weight and Personal Background}
    \label{tb:reporting-error}
\end{table}

\begin{table}[H]
\centering
\begin{tabular}{c|c|c|c|c }
\hline
\hline
\multirow{3}{*}{Variable} & \multicolumn{2}{c|}{Income (Eq. (\ref{eq:Height1}))} & 
    \multicolumn{2}{c}{Income (Eq. (\ref{eq:Height2}))} \\
\cline{2-5}
 & Male & Female & Male & Female \\
\hline
Intercept & \makecell{9.108*** \\ (0.517)} & \makecell{8.747*** \\ (0.510)} 
& \makecell{9.259***\\(0.529)} 
& \makecell{8.666***\\ (0.521)} \\[2ex]

\makecell{Reported Height \\(mm)} & \makecell{4.0e-4* \\ (2.2e-4)} & \makecell{5.3e-4** \\ (2.1e-4)} 
& \makecell{ 2.3e-4\\(2.7e-4)} 
& \makecell{ 6.5e-4***\\(2.2e-4) }\\ [2ex]

\makecell{Reported Weight \\(kg)} & \makecell{ \\} & \makecell{ \\ } 
& \makecell{0.002 \\(0.001)} 
& \makecell{ -0.002\\ (0.001)} \\ [2ex]

Experience & \makecell{0.005 \\ (0.006)} & \makecell{0.017*** \\ (0.006)} 
& \makecell{0.004 \\(0.006)} 
& \makecell{0.018***\\ (0.006)} \\ [2ex]

Experience$^2$ & \makecell{1.0e-4 \\ (1.5e-4)} & \makecell{-4.1e-4*** \\ (1.4e-4)} 
& \makecell{1.0e-4\\(1.5e-4)} 
& \makecell{-4.3e-4***\\ (1.3e-4)}\\ [2ex]

\makecell{Occupation \\(Management)} & \makecell{0.301 \\ (0.253)} & \makecell{0.329 \\ (0.227)} 
& \makecell{0.304\\(0.256)} 
& \makecell{0.327\\ (0.234)}\\
\end{tabular}
    \vspace{0.3cm}
    \caption{The Association between Reported Height/Weight and Family Income}
    \label{tb:reported-height/weight-income}
\end{table}

\begin{table}[H]
\centering
\begin{tabular}{c|c|c|c|c }
\hline
\hline
\multirow{3}{*}{Variable} & \multicolumn{2}{c|}{Income (Eq. (\ref{eq:Height1}))} & 
    \multicolumn{2}{c}{Income (Eq. (\ref{eq:Height2}))} \\
\cline{2-5}
 & Male & Female & Male & Female \\
\hline
Intercept & \makecell{8.608*** \\ (0.509)} & \makecell{8.573*** \\ (0.535)} 
& \makecell{8.775***\\(0.505)} 
& \makecell{8.466***\\ (0.539)} \\[2ex]

\makecell{Height \\(mm)} & \makecell{6.8e-4*** \\ (2.2e-4)} & \makecell{6.4e-4*** \\ (2.4e-4)} 
& \makecell{5.0e-4**\\(2.5e-4)} 
& \makecell{7.9e-4***\\(2.5e-4) }\\ [2ex]

\makecell{Weight \\(kg)} & \makecell{ \\} & \makecell{ \\ } 
& \makecell{0.002 \\(0.001)} 
& \makecell{-0.002*\\ (0.001)} \\ [2ex]

Experience & \makecell{0.005 \\ (0.006)} & \makecell{0.016*** \\ (0.006)} 
& \makecell{0.004 \\(0.006)} 
& \makecell{0.018***\\ (0.006)} \\ [2ex]

Experience$^2$ & \makecell{1.1e-4 \\ (1.5e-4)} & \makecell{-3.9e-4*** \\ (1.3e-4)} 
& \makecell{1.2e-4\\(1.5e-4)} 
& \makecell{-4.1e-4***\\ (1.3e-4)}\\ [2ex]

\makecell{Occupation \\(Management)} & \makecell{0.302 \\ (0.254)} & \makecell{0.328 \\ (0.231)} 
& \makecell{0.305\\(0.252)} 
& \makecell{0.325\\ (0.232)}\\
\end{tabular}
    \vspace{0.3cm}
    \caption{The Association between Height/Weight and Family Income}
     \label{tb:height/weight-income}
\end{table}

\begin{table}[H]
\centering
\begin{tabular}{c|c|c|c|c|c | c  }
\hline
\hline
\multirow{3}{*}{Variable} & \multicolumn{2}{c|}{Income (Eq. (\ref{eq:BMI1}))} & %
    \multicolumn{2}{c|}{Income (Eq. (\ref{eq:BMI2}))}  & %
    \multicolumn{2}{c}{Income (Eq. (\ref{eq:BMI3}))}  \\
\cline{2-7}
 & Male & Female & Male & Female & Male & Female\\
\hline
Intercept & \makecell{9.685*** \\ (0.332)} & \makecell{9.747*** \\ (0.393)} 
& \makecell{9.668*** \\ (0.336)} & \makecell{9.746*** \\(0.389) }
& \makecell{8.965*** \\(0.523)} & \makecell{8.885*** \\ (0.544)} \\[2ex]

Reported BMI & \makecell{0.005 \\(0.004)} & \makecell{-0.005\\ (0.003)} 
& \makecell{-0.007 \\(0.009)} & \makecell{-0.019*** \\(0.007) }
& \makecell{0.005 \\(0.005)} & \makecell{-0.004 \\ (0.003)} \\ [2ex]

\makecell{Reported Height \\(mm)} & \makecell{ \\ } & \makecell{ \\ } 
& \makecell{ \\} & \makecell{ \\ }
& \makecell{3.9e-4* \\(2.1e-4)} & \makecell{5.2e-4** \\ (2.3e-4)}\\ [2ex]

\makecell{Reported Weight \\(kg)} & \makecell{ \\} & \makecell{ \\ } 
& \makecell{0.004* \\(0.002)} & \makecell{0.005** \\ (0.002)}
& \makecell{ \\} & \makecell{ \\ } \\ [2ex]

Experience & \makecell{0.004 \\ (0.006)} & \makecell{0.019*** \\ (0.006)} 
& \makecell{0.004 \\(0.007)} & \makecell{0.018*** \\ (0.006)}
& \makecell{0.005 \\(0.006)} & \makecell{0.018*** \\ (0.005)} \\ [2ex]

Experience$^2$ & \makecell{9.3e-5 \\ (1.5e-4)} & \makecell{-4.5e-4*** \\ (1.3e-4)} 
& \makecell{1.0e-4 \\(1.5e-4)} & \makecell{-4.3e-4*** \\ (1.3e-4)}
& \makecell{1.0e-5 \\(1.5e-4)} & \makecell{-4.3e-4*** \\ (1.2e-4)}\\
\end{tabular}
    \vspace{0.3cm}
    \caption{The Association between Reported BMI and Family Income}
     \label{tb:reported-BMI-income}
\end{table}

\begin{table}[H]
\centering
\begin{tabular}{c|c|c|c|c|c | c  }
\hline
\hline
\multirow{3}{*}{Variable} & \multicolumn{2}{c|}{Income (Eq. (\ref{eq:BMI1}))} & %
    \multicolumn{2}{c|}{Income (Eq. (\ref{eq:BMI2}))}  & %
    \multicolumn{2}{c}{Income (Eq. (\ref{eq:BMI3}))}  \\
\cline{2-7}
 & Male & Female & Male & Female & Male & Female\\
\hline
Intercept & \makecell{9.678*** \\ (0.345)} & \makecell{9.751*** \\ (0.382)} 
& \makecell{9.678*** \\ (0.342)} & \makecell{9.768*** \\(0.384) }
& \makecell{8.499*** \\(0.511)} & \makecell{8.737*** \\ (0.537)} \\[2ex]

BMI & \makecell{ 0.006 \\(0.004)} & \makecell{-0.005*\\ (0.003)} 
& \makecell{-0.018** \\(0.009)} & \makecell{-0.024*** \\(0.007) }
& \makecell{0.005 \\(0.004)} & \makecell{-0.005* \\ (0.003)} \\ [2ex]

\makecell{Height \\(mm)} & \makecell{ \\ } & \makecell{ \\ } 
& \makecell{ \\} & \makecell{ \\ }
& \makecell{6.6e-4*** \\(2.2e-4)} & \makecell{6.3e-4*** \\ (2.4e-4)}\\ [2ex]

\makecell{Weight \\(kg)} & \makecell{ \\} & \makecell{ \\ } 
& \makecell{0.007*** \\(0.002)} & \makecell{0.007** \\ (0.003)}
& \makecell{ \\} & \makecell{ \\ } \\ [2ex]

Experience & \makecell{0.004 \\ (0.006)} & \makecell{0.019*** \\ (0.006)} 
& \makecell{0.004 \\(0.006)} & \makecell{0.018*** \\ (0.006)}
& \makecell{0.004 \\(0.006)} & \makecell{0.018*** \\ (0.005)} \\ [2ex]

Experience$^2$ & \makecell{9.6e-5 \\ (1.5e-4)} & \makecell{-4.4e-4*** \\ ( 1.3e-4)} & \makecell{1.2e-4 \\(1.5e-4)} & \makecell{-4.0e-4*** \\ (1.3e-4)}
& \makecell{1.2e-4 \\(1.5e-4)} & \makecell{-4.1e-4*** \\ (1.3e-4)}\\
\end{tabular}
    \vspace{0.3cm}
    \caption{The Association between BMI and Family Income}
     \label{tb:BMI-income}
\end{table}

\newpage
\begin{longtable}{c| c c}

\hline
\hline\\
        Variable & Male & Female \\[2ex]
\hline\\
\endfirsthead

\hline
\hline\\
        Variable & Male & Female \\[2ex]
\hline\\
\endhead

\hline
\multicolumn{3}{l}{} \\[-10pt] 
    \caption{The Association between Various Body Measures  and Family Income}
     \label{tb:bodymeasures-income}
\endfoot
\multicolumn{3}{l}{} \\[-10pt] 
    \caption{The Association between Various Body Measures  and Family Income (Continued)}
\endlastfoot

Intercept & \makecell{7.173*** \\ (1.428)} & \makecell{9.138*** \\ (1.339)} \\ [2ex]

\makecell{Height \\(mm)} & \makecell{0.001 \\ (0.002)} & \makecell{0.002 \\ (0.002)} \\ [2ex]

\makecell{Weight \\(kg)} & \makecell{-0.003 \\ (0.007)} & \makecell{0.007 \\ (0.007)} \\ [2ex]

\makecell{Acromial Height \\ (Sitting, mm)} & \makecell{0.006*** \\ (0.002)} & \makecell{ \\ } \\ [2ex]

\makecell{Arm Length \\ (Shoulder-to-Elbow, mm)} & \makecell{-0.005** \\ (0.002)} & \makecell{ \\ } \\ [2ex]

\makecell{Chest Circumference \\} & \makecell{0.002* \\ (9.5e-4)} & \makecell{ \\ } \\ [2ex]

\makecell{Buttock \\ (Knee Length, mm)} & \makecell{-0.003** \\ (0.002)} & \makecell{ \\ } \\ [2ex]

\makecell{Elbow Height \\(Sitting, mm)} & \makecell{-0.006*** \\ (0.002)} & \makecell{ \\ } \\
\hline
    
\hline 
\end{longtable}

\begin{table}[H]
\centering
\begin{tabular}{c|c }
\hline
\hline
Variable & $P_3$\\
\hline

\makecell{Bust/Chest Circumference Under Bust} & \makecell{-0.003* \\ (0.002)} \\ [2ex]

\makecell{Chest Circumference \\} & \makecell{0.003** \\ (0.001)} \\ [2ex]

\makecell{Crotch Height \\} & \makecell{-0.005* \\ (0.003)} \\ [2ex]

\makecell{Cup Size \\} & \makecell{-0.066* \\ (0.036)} \\ [2ex]

\makecell{Eye Height, Sitting \\} & \makecell{-0.006* \\ (0.003)} \\ [2ex]

\makecell{Face Length \\} & \makecell{0.019*** \\ (0.005)} \\ [2ex]

\makecell{Hip Breadth, Sitting \\} & \makecell{0.004* \\ (0.002)} \\ [2ex]

\makecell{Hip Circumference Max Height \\} & \makecell{0.003*** \\ (0.001)} \\ [2ex]

\makecell{Subscapular Skinfold \\} & \makecell{-0.010* \\ (0.006)} \\ [2ex]

\makecell{Triceps Skinfold \\} & \makecell{-0.014** \\ (0.006)} \\ [2ex]

\makecell{Waist Front Length  \\} & \makecell{ -0.003** \\ (0.001)} \\ [2ex]

\hline
$\bar{R}^2$ & 0.109 \\ [1ex]
        \makecell[l]{$F$-statistic vs.  constant model } &  4.49 \\ [2ex]
        p-value & 1.2e-18   \\ [2ex]
        N  & 1205  \\
\hline
\end{tabular}
    \vspace{0.3cm}
    \caption{The Association between Female $P_3$ and Various Body Measures}
     \label{tb:bodymeasures-p3}
\end{table}

\begin{table}[H]
\centering
\begin{tabular}{c|c|c|c|c|c | c  }
\hline
\hline
\multirow{3}{*}{Variable} & \multicolumn{2}{c|}{Income (Eq. (\ref{eq:bodyshape1}))} & %
    \multicolumn{2}{c|}{Income (Eq. (\ref{eq:bodyshape2}))}  & %
    \multicolumn{2}{c}{Income (Eq. (\ref{eq:bodyshape3}))}  \\
\cline{2-7}
 & Male & Female & Male & Female & Male & Female\\
\hline
Intercept & \makecell{8.608*** \\ (0.492)} & \makecell{8.573*** \\ (0.534)} 
& \makecell{9.680*** \\ (0.347)} & \makecell{9.751*** \\(0.386) }
& \makecell{8.499*** \\(0.513)} & \makecell{8.490*** \\ (0.625)} \\[2ex]

\makecell{Height \\ (mm)}& \makecell{6.8e-4*** \\(2.1e-4)} & \makecell{6.4e-4***\\ (2.4e-4)} 
& \makecell{ \\} & \makecell{ \\ }
& \makecell{6.6e-4*** \\(2.2e-4)} & \makecell{6.2e-4** \\ (2.4e-4)} \\ [2ex]

BMI & \makecell{ \\ } & \makecell{ \\ } 
& \makecell{0.006 \\ (0.004)} & \makecell{-0.005* \\ (0.003)}
& \makecell{0.005 \\(0.004)} & \makecell{-0.004 \\ (0.003)}\\ [2ex]

\makecell{Hip-to-waist \\Ratio} & \makecell{ \\ } & \makecell{ \\ } 
& \makecell{ \\} & \makecell{ \\ }
& \makecell{ \\} & \makecell{0.002 \\ (0.002)}\\ [2ex]

Experience & \makecell{0.005 \\ (0.006)} & \makecell{0.016*** \\ (0.006)} 
& \makecell{0.004 \\(0.006)} & \makecell{0.019*** \\ (0.006)}
& \makecell{0.004 \\(0.006)} & \makecell{0.018*** \\ (0.006)} \\ [2ex]

Experience$^2$ & \makecell{1.2e-4 \\ (1.6e-4)} & \makecell{-3.9e-4*** \\ (1.3e-4)} 
& \makecell{9.6e-5 \\(1.5e-4)} & \makecell{-4.4e-4*** \\ (1.3e-4)}
& \makecell{1.2e-4 \\(1.5e-4)} & \makecell{-4.0e-4*** \\ (1.3e-4)}\\
\end{tabular}
    \vspace{0.3cm}
    \caption{The Association between BMI/Height/Hip-to-waist-Ratio and Family Income}
     \label{tb:BMI/height/hip2waist-income}
\end{table}

\newpage

\begin{table}[H]
\centering
\begin{tabular}{c|c|c|c|c|c | c  }
\hline
\hline
\multirow{3}{*}{Variable} & \multicolumn{2}{c|}{Income (Eq. (\ref{eq:bodyshape1}))} & %
    \multicolumn{2}{c|}{Income (Eq. (\ref{eq:bodyshape2}))}  & %
    \multicolumn{2}{c}{Income (Eq. (\ref{eq:bodyshape3}))}  \\
\cline{2-7}
 & Male & Female & Male & Female & Male & Female\\
\hline
Intercept & \makecell{9.823*** \\ (0.309)} & \makecell{9.629*** \\ (0.392)} 
& \makecell{9.841*** \\ (0.307)} & \makecell{9.620*** \\(0.392) }
& \makecell{9.823*** \\(0.317)} & \makecell{9.638*** \\ (0.382)} \\[2ex]

$P_1$ & \makecell{0.052*** \\(0.020)} & \makecell{0.033*\\ (0.018)} 
& \makecell{ \\} & \makecell{ \\ }
& \makecell{0.052*** \\(0.019)} & \makecell{0.024 \\ (0.020)} \\ [2ex]

$P_2$ & \makecell{ \\ } & \makecell{ \\ } 
& \makecell{2.0e-4 \\ (0.002)} & \makecell{-0.056*** \\ (0.017)}
& \makecell{-0.002 \\(0.019)} & \makecell{-0.052*** \\ (0.018)}\\ [2ex]

$P_3$ & \makecell{ \\ } & \makecell{ \\ } 
& \makecell{ \\} & \makecell{ \\ }
& \makecell{ \\} & \makecell{0.014 \\ (0.020)}\\ [2ex]

Experience & \makecell{0.004 \\ (0.007)} & \makecell{0.016*** \\ (0.006)} 
& \makecell{0.005 \\(0.007)} & \makecell{0.020*** \\ (0.006)}
& \makecell{0.004 \\(0.006)} & \makecell{0.019*** \\ (0.006)} \\ [2ex]

Experience$^2$ & \makecell{1.1e-4 \\ (1.6e-4)} & \makecell{-4.0e-4*** \\ (1.3e-4)} 
& \makecell{9.5e-5 \\(1.6e-4)} & \makecell{-4.6e-4*** \\ (1.3e-4)}
& \makecell{1.1e-4 \\(1.5e-4)} & \makecell{-4.4e-4*** \\ (1.3e-4)}\\
\end{tabular}
    \vspace{0.3cm}
    \caption{The Association between Body-type Parameters and Family Income}
     \label{tb:bodytype-parameters-income}
\end{table}

\newpage

\begin{longtable}{c|c|c|c|c|c|c}

\hline
\hline
\multirow{3}{*}{Variable} & \multicolumn{2}{c|}{Income (Eq. (\ref{eq:bodyshape3}))} & %
    \multicolumn{2}{c|}{Income (Eq. (\ref{eq:bodyshape3}))}  & %
    \multicolumn{2}{c}{Income (Eq. (\ref{eq:bodyshape3}))}  \\
\cline{2-7}
 & Male & Female & Male & Female & Male & Female\\
\hline

\endfirsthead

\hline
\hline
\multirow{3}{*}{Variable} & \multicolumn{2}{c|}{Income (Eq. (\ref{eq:bodyshape3}))} & %
    \multicolumn{2}{c|}{Income (Eq. (\ref{eq:bodyshape3}))}  & %
    \multicolumn{2}{c}{Income (Eq. (\ref{eq:bodyshape3}))}  \\
\cline{2-7}
 & Male & Female & Male & Female & Male & Female\\
\hline

\endhead

\hline
\multicolumn{3}{l}{} \\[-10pt] 
    \caption{The Association between BMI/Height/Hip-to-waist-Ratio and Family Income - Proxy Variable Approach}
     \label{tb:BMI/Height/Hip-to-waist-Ratio-income-endogeneity-PV-approach}
\endfoot
\multicolumn{3}{l}{} \\[-10pt] 
    \caption{The Association between BMI/Height and Family Income - Proxy Variable Approach (Continued)}
\endlastfoot

Intercept & \makecell{8.482*** \\ (0.511)} & \makecell{8.626*** \\ (0.585)} 
& \makecell{8.510*** \\ (0.513)} & \makecell{8.804*** \\(0.571) }
& \makecell{8.440*** \\(0.508)} & \makecell{8.828*** \\ (0.623)} \\[2ex]

\makecell{Height \\ (mm)} & \makecell{6.5e-4*** \\(2.1e-4)} & \makecell{5.8e-4**\\ (2.4e-4)} 
& \makecell{7.2e-4*** \\ (2.0e-4)} & \makecell{5.2e-4** \\ (2.3e-4)}
& \makecell{7.8e-4*** \\(2.2e-4)} & \makecell{6.1e-4** \\ (2.4e-4)} \\ [2ex]

BMI & \makecell{0.006 \\ (0.004)} & \makecell{-0.005* \\ (0.003)} 
& \makecell{0.005 \\ (0.004)} & \makecell{-0.007** \\ (0.003)}
& \makecell{0.003 \\(0.004)} & \makecell{-0.007** \\ (0.003)}\\ [2ex]

\makecell{Hip-to-waist \\ Ratio} & \makecell{ \\ } & \makecell{0.001 \\ (0.002)} 
& \makecell{ \\ } & \makecell{0.001 \\ (0.002)}
& \makecell{ \\} & \makecell{0.001 \\ (0.002)}\\ [2ex]

Experience & \makecell{0.005 \\ (0.007)} & \makecell{0.018*** \\ (0.006)} 
& \makecell{0.005 \\(0.006)} & \makecell{0.021*** \\ (0.006)}
& \makecell{0.008 \\(0.006)} & \makecell{0.023*** \\ (0.006)} \\ [2ex]

Experience$^2$ & \makecell{9.2e-5 \\ (1.6e-4)} & \makecell{-4.2e-4*** \\ (1.3e-4)} 
& \makecell{9.8e-5 \\(1.5e-4)} & \makecell{-4.7e-4*** \\ (1.3e-4)}
& \makecell{7.6e-5 \\(1.5e-4)} & \makecell{-4.8e-4*** \\ (1.3e-4)}\\ [2ex]

\makecell{Occupation \\(Management)} & \makecell{0.298 \\ (0.245)} & \makecell{0.313 \\ (0.225)} 
& \makecell{0.313 \\(0.250)} & \makecell{0.305 \\(0.214) }
& \makecell{0.236 \\(0.198)} & \makecell{0.243 \\ (0.172)}\\ [2ex]

\makecell{Occupation \\(Blue Collar)} & \makecell{-0.147 \\ (0.270)} & \makecell{-0.118 \\ (0.243)} 
& \makecell{-0.141 \\(0.272)} & \makecell{-0.097 \\ (0.22)}
& \makecell{-0.119 \\(0.216)} & \makecell{-0.089 \\(0.191) }\\
\hline

\end{longtable}

\newpage

\begin{longtable}{c|c|c|c|c|c|c}

\hline
\hline
\multirow{3}{*}{Variable} & \multicolumn{2}{c|}{Income (Eq. (\ref{eq:bodyshape3}))} & %
    \multicolumn{2}{c|}{Income (Eq. (\ref{eq:bodyshape3}))}  & %
    \multicolumn{2}{c}{Income (Eq. (\ref{eq:bodyshape3}))}  \\
\cline{2-7}
 & Male & Female & Male & Female & Male & Female\\
\hline

\endfirsthead

\hline
\hline
\multirow{3}{*}{Variable} & \multicolumn{2}{c|}{Income (Eq. (\ref{eq:bodyshape3}))} & %
    \multicolumn{2}{c|}{Income (Eq. (\ref{eq:bodyshape3}))}  & %
    \multicolumn{2}{c}{Income (Eq. (\ref{eq:bodyshape3}))}  \\
\cline{2-7}
 & Male & Female & Male & Female & Male & Female\\
\hline

\endhead

\hline
\multicolumn{3}{l}{} \\[-10pt] 
    \caption{The Association between Body-type Parameters and Family Income - Proxy Variable Approach}
     \label{tb:bodytype-parameters-income-endogeneity-PV-approach}
\endfoot
\multicolumn{3}{l}{} \\[-10pt] 
    \caption{The Association between Body-type Parameters and Family Income - Proxy Variable Approach (Continued)}
\endlastfoot

Intercept & \makecell{9.819*** \\ (0.315)} & \makecell{9.638*** \\ (0.403)} 
& \makecell{9.937*** \\ (0.303)} & \makecell{9.669*** \\(0.367) }
& \makecell{9.933*** \\(0.326)} & \makecell{9.784*** \\ (0.390)} \\[2ex]

$P_1$ & \makecell{0.054*** \\(0.019)} & \makecell{ 0.020\\ (0.020)} 
& \makecell{0.054*** \\ (0.019)} & \makecell{ 0.016 \\ (0.019)}
& \makecell{0.056*** \\(0.018)} & \makecell{0.027 \\ (0.019)} \\ [2ex]

$P_2$ & \makecell{0.002 \\ (0.018)} & \makecell{-0.059*** \\ (0.017)} 
& \makecell{-0.007 \\ (0.019)} & \makecell{-0.070*** \\ (0.018)}
& \makecell{-0.015 \\(0.018)} & \makecell{-0.064*** \\ (0.017)}\\ [2ex]

$P_3$ & \makecell{ \\ } & \makecell{ 0.005 \\ (0.020)} 
& \makecell{ \\} & \makecell{0.004 \\ (0.020)}
& \makecell{ \\} & \makecell{0.003 \\ (0.018)}\\ [2ex]

Experience & \makecell{0.005 \\ (0.006)} & \makecell{0.020*** \\ (0.006)} 
& \makecell{0.005 \\(0.006)} & \makecell{0.023*** \\ (0.006)}
& \makecell{0.008 \\(0.006)} & \makecell{0.024*** \\ (0.006)} \\ [2ex]

Experience$^2$ & \makecell{8.4e-5 \\ (1.5e-4)} & \makecell{-4.5e-4*** \\ (1.3e-4)} 
& \makecell{9.0e-5 \\(1.5e-4)} & \makecell{-5.0e-4*** \\ (1.3e-5)}
& \makecell{6.9e-5 \\(1.5e-4)} & \makecell{-5.0e-4*** \\ (1.4e-4)}\\ [2ex]

\makecell{Occupation \\(Management)} & \makecell{0.297 \\ (0.247)} & \makecell{0.318 \\ (0.233)} 
& \makecell{0.311 \\(0.255)} & \makecell{0.309 \\(0.208) }
& \makecell{0.234 \\(0.199)} & \makecell{0.249 \\ (0.178)}\\ [2ex]

\makecell{Occupation \\(Blue Collar)} & \makecell{-0.148 \\ (0.272)} & \makecell{-0.122 \\ (0.248)} 
& \makecell{-0.143 \\(0.275)} & \makecell{-0.099 \\ (0.227)}
& \makecell{-0.120 \\(0.209)} & \makecell{-0.089 \\(0.191) }\\
\hline

\end{longtable}

\newpage

\begin{longtable}{c|c|c|c|c}

\hline
\hline
\multirow{3}{*}{Variable} & \multicolumn{2}{c|}{Height ($1^{\text{st}}$-step Eq. (\ref{eq:cf-1st-stage-equation1}))} & %
    \multicolumn{2}{c}{Income ($2^{\text{nd}}$-step Eq. (\ref{eq:cf-2nd-stage-equation}))}  \\
\cline{2-5}
 & Male & Female & Male & Female\\
\hline

\endfirsthead

\hline
\hline
\multirow{3}{*}{Variable} & \multicolumn{2}{c|}{Height ($1^{\text{st}}$-step Eq. (\ref{eq:cf-1st-stage-equation1}))} & %
    \multicolumn{2}{c}{Income ($2^{\text{nd}}$-step Eq. (\ref{eq:cf-2nd-stage-equation}))}  \\
\cline{2-5}
 & Male & Female  & Male & Female\\
\hline

\endhead

\hline
\multicolumn{3}{l}{} \\[-10pt] 
         \caption{The Association between BMI/Height/Hip-to-waist-Ratio and Family Income - Control Function Approach}
     \label{tb:BMI/Height/Hip-to-waist-Ratio-income-endogeneity-CF-approach}
     
\endfoot
\multicolumn{3}{l}{} \\[-10pt] 
    \caption{The Association between BMI/Height and Family Income - Control Function Approach (Continued)}
\endlastfoot

Intercept & \makecell{1832.318*** \\ (32.212)} & \makecell{1634.629*** \\ (27.136)} 
& \makecell{7.785*** \\(1.104)} & \makecell{10.674*** \\ (1.294)} \\[2ex]

\makecell{Height \\ (mm)} & \makecell{ \\} & \makecell{ \\ } 
& \makecell{0.001* \\(5.9e-4)} & \makecell{-6.4e-4 \\ (7.5e-4)} \\ [2ex]

BMI & \makecell{ \\ } & \makecell{ \\ } 
& \makecell{0.004 \\(0.004)} & \makecell{-0.005 \\ (0.003)}\\ [2ex]

\makecell{Hip-to-waist \\ Ratio} & \makecell{ \\ } & \makecell{ \\ } 
& \makecell{ \\} & \makecell{0.002  \\ (0.002)}\\ [2ex]

$\hat{\nu}$ & \makecell{ \\ } & \makecell{  \\ } 
& \makecell{-6.0e-4 \\ (6.5e-4)} & \makecell{0.001* \\ (8.0e-4)}\\ [2ex]

\makecell{Shoe Size \\} & \makecell{13.502*** \\ (4.958)} & \makecell{19.847***\\ (4.187)} 
& \makecell{ \\} & \makecell{  \\ }\\ [2ex]

\makecell{Jacket Size \\ (Blouse Size)} & \makecell{10.132*** \\ (1.309)} & \makecell{7.425*** \\ (1.874)} 
& \makecell{ \\} & \makecell{  \\ } \\[2ex]

\makecell{Pants Size \\ } & \makecell{8.372*** \\ (2.384)} & \makecell{1.010 \\ (1.416)} 
& \makecell{} & \makecell{  \\ } \\
\hline

\end{longtable}

\newpage

\begin{longtable}{c|c|c|c|c}

\hline
\hline
\multirow{3}{*}{Variable} & \multicolumn{2}{c|}{$P_1$ ($1^{\text{st}}$-step Eq. (\ref{eq:cf-1st-stage-equation1}))} & %
    \multicolumn{2}{c}{Income ($2^{\text{nd}}$-step Eq. (\ref{eq:cf-2nd-stage-equation}))}  \\
\cline{2-5}
 & Male & Female  & Male & Female\\
\hline

\endfirsthead

\hline
\hline
\multirow{3}{*}{Variable} & \multicolumn{2}{c|}{$P_1$ ($1^{\text{st}}$-step Eq. (\ref{eq:cf-1st-stage-equation1}))} & %
    \multicolumn{2}{c}{Income ($2^{\text{nd}}$-step Eq. (\ref{eq:cf-2nd-stage-equation}))}  \\
\cline{2-5}
 & Male & Female  & Male & Female\\
\hline

\endhead

\hline
\multicolumn{3}{l}{} \\[-10pt] 
    \caption{The Association between Body-type Parameters and Family Income - Control Function Approach}
     \label{tb:bodytype-parameters-income-endogeneity-CF-approach}
\endfoot
\multicolumn{3}{l}{} \\[-10pt] 
    \caption{The Association between Body-type Parameters and Family Income - Control Function Approach (Continued)}
\endlastfoot

Intercept & \makecell{0.546 \\ (0.410)} & \makecell{-0.181 \\ (0.347)} 
& \makecell{9.945*** \\(0.336)} & \makecell{9.735*** \\ (0.365)} \\[2ex]

$P_1$ & \makecell{ \\} & \makecell{ \\ } 
& \makecell{0.097** \\(0.049)} & \makecell{-0.074 \\ (0.056)} \\ [2ex]

$P_2$ & \makecell{ \\ } & \makecell{ \\ } 
& \makecell{6.5e-4 \\(0.019)} & \makecell{-0.069*** \\ (0.018)}\\ [2ex]

$P_3$ & \makecell{ \\ } & \makecell{  \\ } 
& \makecell{ \\} & \makecell{0.009 \\ (0.020)}\\ [2ex]

$\hat{\nu}$ & \makecell{ \\ } & \makecell{  \\ } 
& \makecell{-0.058 \\ (0.055)} & \makecell{0.101* \\ (0.058)}\\ [2ex]

\makecell{Shoe Size \\} & \makecell{0.159** \\ (0.062)} & \makecell{0.193***\\ (0.053)} 
& \makecell{ \\} & \makecell{  \\ }\\ [2ex]

\makecell{Jacket Size \\ (Blouse Size)} & \makecell{0.118*** \\ (0.016)} & \makecell{0.099** \\ (0.025)} 
& \makecell{ \\} & \makecell{  \\ } \\[2ex]

\makecell{Pants Size \\} & \makecell{0.123*** \\ (0.030)} & \makecell{0.040** \\ (0.020)} 
& \makecell{ \\} & \makecell{  \\ } \\
\hline

\end{longtable}

\end{document}